\documentclass[prd,aps
,tightenlines,
superscriptaddress,showpacs,nofootinbib,eqsecnum,amsfonts,amsmath,floatfix
]{revtex4}
\usepackage{latexsym}
\usepackage{amssymb}
\usepackage{amsfonts}
\usepackage{amsmath}
\usepackage{bm}
\usepackage[dvips]{graphicx}
\usepackage[usenames]{color}
\usepackage{ulem}
\usepackage{multirow}

\def\be{\begin{equation}}
\def\ee{\end{equation}}

\def\gsim{\mathrel{
\rlap{\raise 0.511ex \hbox{$>$}}{\lower 0.511ex
\hbox{$\sim$}}}}
\def\lsim{\mathrel{
\rlap{\raise 0.511ex \hbox{$<$}}{\lower 0.511ex
\hbox{$\sim$}}}}

\begin{document}
\title{Comparison of post-Newtonian templates for  compact binary 
 inspiral signals in gravitational-wave detectors}

\author{Alessandra Buonanno}\email{buonanno@umd.edu}
\affiliation{Maryland Center for Fundamental Physics, 
Department of Physics, University of Maryland, College Park, MD 20742, USA }
\author{Bala R Iyer}\email{bri@rri.res.in}
\affiliation{Raman Research Institute, Bangalore, 560 080, India}
\affiliation{School of Physics and Astronomy, Cardiff University, 
5, The Parade, Cardiff, UK, CF24 3YB}
\author{Evan Ochsner}\email{evano@umd.edu}
\affiliation{Maryland Center for Fundamental Physics, 
Department of Physics, University of Maryland, College Park, MD 20742, USA }
\author{Yi Pan}\email{ypan@umd.edu}
\affiliation{Maryland Center for Fundamental Physics, 
Department of Physics, University of Maryland, College Park, MD 20742, USA }
\author{B S Sathyaprakash}
\email{B.Sathyaprakash@astro.cf.ac.uk}
\affiliation{School of Physics and Astronomy, Cardiff University, 
5, The Parade, Cardiff, UK, CF24 3YB}
\date{\today}
\begin{abstract}
The two-body dynamics in general relativity has been solved
perturbatively using the post-Newtonian (PN) approximation. The
evolution of the orbital phase and the emitted gravitational
radiation are now known to a rather high order up to ${\cal O}(v^8)$, 
$v$ being the characteristic velocity of the binary. The orbital 
evolution, however, cannot be specified uniquely due 
to the inherent freedom in the choice of parameter used in 
the PN expansion as well as the method 
pursued in solving the relevant differential equations. The goal 
of this paper is to determine the (dis)agreement between different 
PN waveform families in the context of initial and advanced 
gravitational-wave detectors. The waveforms employed in our analysis 
are those that are currently used by Initial LIGO/Virgo, that is 
the time-domain PN models TaylorT1, TaylorT2, TaylorT3, 
the Fourier-domain representation TaylorF2 (or stationary phase 
approximant, SPA) and the effective-one-body (EOB) model, and two 
more recent models, TaylorT4 and TaylorEt.
For these models we examine their overlaps with one 
another for a number of different binaries at 2PN, 3PN and 3.5PN
orders to quantify their 
differences.  We then study the overlaps of these families with the  
prototype effective-one-body family, currently used by Initial LIGO, calibrated 
to numerical relativity simulations to help us decide whether there exist  
preferred families, in terms of  detectability and computational cost,
that are the most appropriate as search templates. We conclude that as long 
as the total mass remains less than a certain upper limit 
$M_{\rm crit}$, all template families at 3.5PN order (except TaylorT3 and TaylorEt) 
are equally good for the purpose of detection. The value of $M_{\rm crit}$ is found to be 
$\sim 12\,M_\odot$ for Initial, Enhanced and Advanced LIGO. From a purely computational point of view
we recommend that 3.5PN TaylorF2 be used below $M_{\rm crit}$ and EOB 
calibrated to numerical relativity simulations be used for  total binary
mass $M > M_{\rm crit}$.
\end{abstract}
\maketitle

\section{Introduction} 
\label{sec:intro}

Sensitivity of several interferometric gravitational-wave detectors 
has either already reached, or is close to, the design goals that 
were set more than a decade ago~\cite{bss:ligo,Abbott:2003vs,bss:virgo,
Acernese:2006bj,Luck:2006ug,1997gwd..conf..183T,HOUGHREF}. Upgrades that are currently
underway and planned for the next four to five years, will
see their sensitivity improve by factors of a few to an 
order-of-magnitude~\cite{Smith:2009}. Coalescing binaries consisting of neutron
stars and/or black holes are probably the most promising
sources for a first direct detection of gravitational waves.
At current sensitivities, initial interferometers are capable of 
detecting binary neutron star inspirals at distances
up to 30 Mpc, the range increasing to 60 Mpc for enhanced
detectors (circa 2009-2011) and 175 Mpc for advanced detectors
(circa 2014+). Binary black holes or a mixed system consisting
of a neutron star and a black hole can be detected to a far
greater distance depending on the total mass and the mass ratio.

The range of interferometric detectors for coalescing binaries
is computed by assuming that one can pull the signal out of noise
by matched filtering. This in turn means that one is able to 
follow the phasing of gravitational waves typically to within
a fraction of a cycle over the duration of the signal in band. 
The reason for this optimism comes from the fact that one knows the phase
evolution of the signal to a high order in post-Newtonian (PN) formalism
~\cite{BLANCHETREF}.
Several authors have assessed whether the accuracy with which
the formalism provides the waveforms is good enough for the purpose 
of detection and parameter estimation~\cite{Cutler:1993vq,Cutler:1994ys,PoissonAndWill,Tanaka:1997dj,
Damour1998,Droz:1999qx,Damour:2000gg,DIS01,Damour:2002kr,Damour:2002vi,Canitrot:2001,
BCV1,BCV2,Arun:2004hn,Ajith:2004ys,Lindblom:2008cm}. 
The problem, as we shall see below, is complicated  since the PN approximation does not lead 
to a unique model of the phase evolution.  Moreover, though PN 
results are good up to mildly relativistic velocities, the standard 
PN approximants become less and less accurate 
in the strongly relativistic regime as one approaches the last stable orbit 
(LSO). Resummation methods~\cite{Damour1998} and
in particular the effective-one-body (EOB)~\cite{BuonD98,Buonanno:2000ef,
Damour:2000we} extensions of the PN
approximants, are needed for analytical treatments close to and beyond
the LSO.

The success in numerical-relativity simulations  of binary black holes
~\cite{Pretorius05,Campanelli:2005dd,Baker:2005vv,Boyle:2007ft,Pretorius:2007nq} 
now provides results for gravitational waveforms that can be  compared
to standard PN results and other resummed extensions.  On the one hand, 
the analytical PN results for the inspiral phase of the evolution
are needed to calibrate and interpret the numerical relativity
waveforms of coalescence and merger. On the other hand, the numerical
relativity results extend the analytical approximations beyond the 
inspiral phase and provide the important coalescence and merger phases, 
producing the strongest signals that are crucial for the detection of 
binary black holes. However, numerical simulations are still 
computationally expensive and time-consuming and presently only a small 
region of the parameter space can be explored. Even in the foreseeable future,
numerical relativity may not be able to handle,
tens of thousands of cycles that are expected from highly asymmetric 
systems (e.g., a neutron star falling into an intermediate-mass 
black hole of 100 $M_\odot$) or low-mass symmetric systems (e.g., a 
binary neutron star).  
Analytical models that smoothly go from the inspiral through coalescense to
quasi-normal ringing would be needed and this has led to
phenomenological templates~\cite{Ajith:2007kx,Pan:2007nw,Boyle:2009dg}. 
and EOB waveforms~\cite{Buonanno-Cook-Pretorius:2007,Pan:2007nw,Damour:2007xr,Buonanno:2007pf,Damour:2007yf,
Damour:2007vq,Damour:2008te,Boyle2008a,Damour:2009b,Buonanno:2009qa}. In particular, the recent improved 
EOB models~\cite{Damour:2009b,Buonanno:2009qa} which 
also incorporate a multiplicative decomposition of 
the multipolar waveform into several physically motivated factors
supplemented by a suitable hybridisation (using test particle results)
~\cite{Damour:2009a}, and an improved treatment of  non-quasi-circular corrections,
show evidence of remarkable success in modeling accurately the  
numerical relativity waveforms for different mass ratios. 

The emphasis of this work is different. Recently, there have been 
investigations 
~\cite{Bose:2008ix} 
on the ability of various standard 
families of PN templates to  detect a specific signal model 
TaylorEt~\cite{Gopakumar:2007jz,Gopakumar:2007vh,Tessmer:2008tq}
and the often-used TaylorF2 to detect a complete
numerical relativity signal including merger and ringdown~\cite{Pan:2007nw,Boyle:2009dg}.
Reference~\cite{Bose:2008ix} modelled the signal by the TaylorEt approximant
at 3.5PN order and looked at the effectualness and systematic biases
in the estimation of mass parameters for 
TaylorT1, TaylorT4 and TaylorF2 templates in the LIGO and Virgo detectors.
It also looked into the possibility of improving the effectualness
by using unphysical values of $\nu$ beyond the maximum value of $0.25$.
It was found that the overlaps of a TaylorEt signal with TaylorT1, TaylorT4
and TaylorF2 template is smaller than $0.97$ and involved for equal-mass
systems a large bias in the  total mass. For unequal-mass systems
higher overlaps can be obtained at the cost of a large bias in
mass and symmetric mass ratio $\nu$ and which can be
further improved by unphysical values of $\nu >0.25$.  The templates are more unfaithful with
increasing total mass. To detect optimally the complete numerical-relativity signal, 
including merger and ringdown, Ref.~\cite{Pan:2007nw} suggested 
the possibility of using the TaylorF2 template bank with a 
frequency cutoff $f_c$ larger than the usual upper cutoff (i.e., the Schwarzschild LSO) and 
closer to the fundamental quasi-normal mode frequency of the final black hole.   
Moreover, they proposed to further improve this family by  allowing either for unphysical values of $\nu$ or 
for the inclusion of a pseudo 4PN (p4PN) coefficient in the template phase, calibrated to the 
numerical simulations. 
Reference~\cite{Boyle:2009dg} extended the results of Ref.~\cite{Pan:2007nw}  
to more accurate numerical waveforms, found that 3.5PN templates 
 {\it are nearly 
always better and rarely significantly worse} than the 2PN templates, and proposed simple analytical frequency 
cutoffs for both Initial and Advanced LIGO --- for example for Initial LIGO 
they recommended a strategy using p4PN templates for $M \leq 35 M_\odot$ and 
3.5PN templates with unphysical values of $\nu$ for larger masses. However, we 
notice that there is no reason for changing the template bank above $35 M_\odot$. 
Reference~\cite{Boyle:2009dg} could have used the p4PN templates over the entire mass region, 
if they had not employed in their analysis the p4PN coefficient used in Ref.~\cite{Pan:2007nw}, but had calibrated it to the highly 
accurate waveforms used in their paper\footnote{We computed that the
p4PN coefficient calibrated to the highly accurate waveforms used in
Ref.~\cite{Boyle:2009dg} is ${\mathcal Y}=3714$, instead of ${\mathcal
Y}=3923$ found in Ref.~\cite{Pan:2007nw}.}. 

In this work our primary focus is on binary systems dominated
by early inspiral and on a critical study of the variety of approximants
that describe this. Towards this end, in this paper we  will provide 
a sufficiently exhaustive comparison of different PN  models of adiabatic inspiral
for an illustrative variety of different systems 
and quantify how (dis)similar they are for the purpose of 
detection. The choice of the PN models used in this paper is motivated 
by the fact that they are available in the {\it LIGO Algorithms Library} (LAL) and  
some of them have been used in the searches by Initial LIGO.  
We also compare all these PN models with one fiducial EOB model calibrated to numerical-relativity simulations~\cite{Buonanno:2007pf} 
to delineate the range of mass values where one must definitely 
go beyond the inspiral-dominated PN models to a more complete description
including plunge and coalescence. The choice of this fiducial, preliminary  
EOB model is only motivated by the fact that it is the EOB model available in 
LAL and it is currently used for searches by Initial LIGO. It will be improved in the future using the recent results 
in Refs.~\cite{Damour:2009b,Buonanno:2009qa}. We will conclude that for total masses below  a certain upper limit 
$M_{\rm crit},$ all template families at 3.5PN order
(except for TaylorT3 and TaylorEt) are equally good for the 
purpose of detection. 
 $M_{\rm crit}$ is found to be $\sim 12\,M_\odot$ for Initial, Enhanced and Advanced LIGO.
Based  solely on computational costs,
we recommend that 3.5PN  TaylorF2 be used below $M_{\rm crit}$ and EOB calibrated to numerical relativity 
simulations be used for total binary mass $M > M_{\rm crit}$.

The paper is organized as follows. In Sec.~\ref{sec:status} we summarise the present status of the PN 
approximation. In Sec.~\ref{sec:PNA}
we recapitulate for completeness the main PN approximants and 
try to provide a ready-reckoner for the equations
describing them and the relevant initial and termination conditions. 
In Sec.~\ref{sec:freq evolution} we discuss  the frequency evolution
in each of these models.  In Sec.~\ref{sec:overlaps} we discuss overlaps and the 
maximization used in this work. Section~\ref{sec:results} and \ref{sec:region} 
presents the results of our analysis related to the effectualness, while Sec.~\ref{sec:faithfulness} 
summarizes the results related to the faithfulness. In Sec.~\ref{sec:conclusions} we summarize 
our main conclusions. Readers who are interested in the main results of the paper and want 
to avoid technical details could skip Secs.~\ref{sec:status}, \ref{sec:PNA}, \ref{sec:freq evolution}, 
and \ref{sec:overlaps}, read the main results of Secs.~\ref{sec:results}, \ref{sec:region} and \ref{sec:faithfulness}, 
and mainly focus on Sec.~\ref{sec:conclusions}.

\section{Current status of post-Newtonian approximation} \label{sec:status}

Post-Newtonian approximation computes the evolution of the
orbital phase $\phi(t)$ of a compact binary as a perturbative expansion in a small parameter,
typically taken as $v=(\pi M F)^{1/3}$ (characteristic velocity in the binary), or 
$x=v^2,$ although other variants exist.
Here $M$ is the total mass of the binary and $F$ the gravitational-wave frequency.
In the adiabatic approximation,
and for the {\it restricted} waveform in which case the gravitational
wave phase is twice the orbital phase, the theory allows the phasing 
to be specified by a pair of differential equations
 $\dot\phi(t) =  v^3/ M,$ $\dot v = -{\cal{F}}(v)/E'(v),$ where $M$ is the total mass
of the system, ${\cal{F}}$ its gravitational-wave luminosity and $E'(v)$ is
the derivative of the binding energy with respect to $v.$ Different
PN families arise because one can choose to treat
the ratio ${\cal F}/E'(v)$ differently while being equivalent with the same
PN order~\cite{DIS01}. For instance, one can leave the PN expansions of the luminosity
${\cal F}(v)$ and $E'(v)$ as they appear (the so-called { TaylorT1} model), 
or expand the rational polynomial ${\cal F}(v)/E'(v)$ in $v$ 
 to consistent PN order (the {TaylorT4} model), recast as a pair of parametric equations $\phi(v)$ and $t(v)$ 
(the { TaylorT2} model), or the phasing could be written as an
explicit function of time $\phi(t)$ (the { TaylorT3 model}). 
These different representations are made possible because one is 
dealing with a perturbative series.  Therefore, one is at liberty 
to ``resum'' or ``reexpand'' the series in any way one wishes (as long 
as one keeps terms to the correct order in the perturbation expansions),
or even retain the expression as the quotient of two polynomials
and treat them numerically. There is also the freedom of writing the series in a different 
variable, say (suitably adimensional) $E$ (the so called { TaylorEt} model).

In addition to these models, there have been efforts to extend the
evolution of a binary beyond what is naturally prescribed by the
PN formalism. Let us briefly discuss two reasons why
the PN evolution cannot be used all the way up to the merger of the
two bodies. PN evolution is based on the so-called adiabatic
approximation according to which the fractional change in the 
orbital frequency $F_{\rm orb}$ over each orbital period is negligibly small,
i.e. $\dot F_{\rm orb}/F_{\rm orb}^2 \ll 1$.  This 
assumption is valid during most of the evolution, but begins to 
fail as the system approaches the LSO where
$f_{\rm LSO}=( 6^{3/2} \pi M)^{-1}$. In some cases, 
the frequency evolution stops from being monotonic and $\dot f$ changes
from being positive to negative well before reaching the LSO ---
an indication of the breakdown of the approximation. 

From the view point of maximizing detection potential one is also
interested in going beyond the inspiral phase. The merger and
ringdown phases of the evolution, when the luminosity is greatest,
cannot be modelled by standard PN approximation. The use of  
resummation techniques more than a decade ago was followed by the
construction of the EOB model~\cite{BuonD98,Buonanno:2000ef,Damour:2000we}, which
has analytically provided the plunge, merger and ringdown phases of 
the binary evolution.  As mentioned before, more recently, 
these models have been calibrated to numerical relativity simulations~
\cite{Buonanno-Cook-Pretorius:2007,Pan:2007nw,Damour:2007xr,Buonanno:2007pf,Damour:2007yf,
Damour:2007vq,Damour:2008te,Boyle2008a,Damour:2009b,Buonanno:2009qa}.
We now have a very reliable EOB model that can be used to model the merger dynamics.
 
An astronomical binary is characterized by a large number of parameters
some of which are intrinsic to the system (e.g., the masses and spins 
of the component stars and the changing eccentricity of the orbit)
and others that are extrinsic (e.g., source location and orientation 
relative to the detector). In this paper we will worry about only the
detection problem. Furthermore, we will assume that a coincident detection
strategy will be followed so that we do not have to worry about the
angular parameters such as the direction to the source, wave's polarization,
etc. If binaries start their lives when their separation $r$ is 
far larger compared to their gravitational radius (i.e., $r\gg GM/c^2$),
by the time they enter the sensitivity band of ground-based detectors
any initial eccentricity would have been lost due to gravitational
radiation reaction, which tends to circularize\footnote{Though this 
assumption is justified for the prototypical binaries we focus on in this 
work, there exist credible astrophysical scenarios 
that lead to inspiral signals from binaries with non-negligible eccentricity
in the sensitive detector bandwidth. A more involved treatment is then called
for and available. 
See e.g.~\cite{MartPois99,DGI04,KG06,Tessmer:2007jg,ABIQ07tail,ABIQ07,Yunes:2009yz}.}
a binary~\cite{Peters:1963ux,Pe64}. Therefore,
we shall consider only systems that are on a quasi-circular 
inspiralling orbit. We shall also neglect spins which means that we
have to worry in reality about only the two masses of the component
bodies.

Our goal is to explore how (dis)similar the different
waveform families are. We do this by computing the (normalized)
cross-correlation between signals and templates, maximized either 
only over the extrinsic parameters of the templates ({\it faithfulness}) 
or over the intrinsic and extrinsic parameters of the templates 
({\it effectualness}), the noise power spectral density of the detector serving as 
a weighting factor in the computation of the correlation 
(see Sec.\ref{sec:overlaps}). Our conclusions, therefore, will 
depend on the masses of the compact stars as well as the 
detector that we hope to observe the signal with.

The overlaps (i.e., the normalized cross-correlation maximized over
various parameters and weighted by the noise power spectral density) 
we shall compute are sensitive to the shape of the noise spectral density 
of a detector and not on how deep that sensitivity is.  
Now, the upgrade from initial to advanced interferometers will see 
improvements in sensitivity not only at a given frequency but over a 
larger band. Therefore, the agreement between different PN models will
be sensitive to the noise spectral density that is used in the inner product. 
Thus, we will compare the PN families using power spectral densities of initial
and advanced interferometric detectors. 

We end this brief overview
 with the following observation. As mentioned earlier, following
all present gravitational wave data analysis pipelines, this paper works
only in the restricted wave approximation. This approximation assumes the 
waveform amplitude to be Newtonian and thus includes only the 
leading second harmonic of the orbital phase. 
Higher PN order amplitude terms bring in harmonics 
of the orbital phase other than the dominant one at twice the orbital
frequency. Their effects can be significant
~ \cite{VanDenBroeck:2006ar,VanDenBroeck:2006qu}, especially close to 
merger~\cite{Damour:2009b}, 
and they need to be carefully included in future work.

\section{The PN Approximants} \label{sec:PNA}

For the convenience of the reader, in this section, we recapitulate
the basic formulas for the different PN families  
from Refs.~\cite{DIS01,Damour:2002kr}.
While comparing the expressions below
to those in Refs.~\cite{DIS01,Damour:2002kr} recall 
$\lambda=-1987/3080$~\cite{Damour:2001bu,Blanchet:2003gy}
and $\Theta=-11831/9240$~\cite{Blanchet:2004ek,Blanchet:2005tk}.  
In addition to the evolution equations, we 
shall also provide initial and final conditions.  From the perspective 
of a data analyst, the initial condition is simply a starting frequency 
$F_0$ and phase $\phi_0,$ which can be translated,
with the help of evolution equations, as conditions on the relevant
variables.  We shall also give explicit expressions for
the evolution of the gravitational wave 
frequency, namely $\dot F \equiv dF/dt,$ or
more precisely, the dimensionless quantity $\dot F\,F^{-2},$ 
in Sec.\ \ref{sec:freq evolution}, where they will be used to
study the rate at which the binary coalesces in different PN families,
which will help us understand the qualitative difference between
them.  The contents of this section should act as a single point of 
resource for anyone who is interested in implementing the 
waveforms for the purpose of data analysis and other applications.  

The basic inputs for all families are the PN expressions for
the conserved 3PN energy (per unit total mass) 
~\cite{Damour:2001bu,Blanchet:2003gy,Damour:2000ni,
deAndrade:2000gf, Blanchet:2002mb,Itoh:2003fy}
 $E_3(v)$ and
 3.5PN energy flux 
~\cite{Blanchet:2001aw,Blanchet:2001ax,Blanchet:2004bb,
 Blanchet:2004ek,Blanchet:2005tk}  
 ${\cal F}_{3.5}(v)$, 
\begin{widetext}
\begin{eqnarray}
E_3(v)& = & -\frac{1}{2} \nu v^2   \left[ 1 - 
\left (\frac{3}{4} + \frac{1}{12} \nu \right ) v^2 
- \left( \frac{27}{8} -\frac{19}{8} \nu  + \frac{1}{24}\nu^2 \right) v^4 
      - \left\{ \frac{675}{64}
- \left(\frac{34445}{576}-\frac{205 }{96}\pi^2\right)\nu 
+\frac{155 }{96}\nu^2
+\frac{35 }{5184}\nu^3
\right\} v^6 \right],
\label{eq:energy}
\nonumber \\ \\
{\cal{F}}_{3.5}(v) & = &\frac{32}{5} \nu^2 v^{10}  \left[1 
     - \left( \frac{1247}{336}+ \frac{35  }{12}\nu \right) v^2
     + 4 \pi  v^3 - \left( \frac{44711}{9072}
-\frac{9271  }{504}\nu- \frac{65 }{18}\nu^2 \right) v^4 
     - \left( \frac{8191}{672} +\frac{583  }{24}\nu \right)\pi  v^5 \right.\nonumber\\
     & + & \left\{ 
      \frac{6643739519}{69854400} + \frac{16 }{3} \pi^2
- \frac{1712 }{105} \gamma
     + \left(\frac{41 }{48}\pi^2-\frac{134543}{7776}\right) \nu 
     - \frac{94403 }{3024} \nu^2
-\frac{775 }{324}\nu^3
-\frac{856}{105} \log \left(16 v^2\right) \right \} v^6 \nonumber \\
     & -  & \left. \left(
      \frac{16285}{504}
-\frac{214745 }{1728}\nu
-\frac{193385 }{3024}\nu^2 \right)\pi v^7 \right],
\label{eq:flux}
\end{eqnarray}
\end{widetext}
where $\gamma=0.577216\ldots$ is the Euler constant.
In the adiabatic approximation one assumes that the orbit evolves 
slowly so that the fractional change in the orbital velocity 
$\omega$ over an orbital period is negligibly small. That is, 
$\frac{\Delta\omega}{\omega} \ll 1,$ or, 
equivalently, $\frac{\dot \omega}{\omega^2} \ll 1.$ In this 
approximation, one expects the luminosity in gravitational waves
to come from the change in orbital energy averaged over a period.
For circular orbits this means one can use the energy balance
equation ${\cal F}= -d{\cal E}/dt$ where ${\cal E}= M E$. 

In the adiabatic approximation one can write an equation for the
evolution of any of the binary parameters.  For instance, the 
evolution of the orbital separation $r(t)$ can be
written as 
$\dot r(t) = \dot {\cal E} / (d{\cal E}/dr) = -{\cal F}/(d{\cal E}/dr).$
Together with the Kepler's law, the energy balance equation can
be used to obtain the evolution of the orbital phase\footnote{
Recall that the gravitational-wave  phase is
twice the orbital phase for the restricted waveform and leads to differences
in factors of $2$ between the equations here for the
 {\it orbital} phase and
those in~\cite{DIS01} for the {\it gravitational-wave} phase.}:
\begin{subequations}
\begin{eqnarray}
\label{eq:phasing formula general2a}
\frac{d\phi}{dt} - \frac{v^3}{M} & = & 0,\\
\frac{dv}{dt}  + \frac{{\cal F}(v)}{ME'(v)} & = & 0,
\label{eq:phasing formula general2b}
\end{eqnarray}
\label {eq:phasing formula general2}
\end{subequations}
or, equivalently,
\begin{subequations}
\begin{eqnarray}
t(v) & = & t_{\rm ref} + M \int_v^{v_{\rm ref}} dv \,
\frac{E'(v)}{{\cal F}(v)}, \\
\phi (v) & = & \phi_{\rm ref} +  \int_v^{v_{\rm ref}} dv\, v^3 \,
\frac{E'(v)}{{\cal F}(v)},
\end{eqnarray}
\label {eq:phasing formula general1}
\end{subequations}
where $t_{\rm ref}$ and $\phi_{\rm ref}$ are integration constants and
$v_{\rm ref}$ is an arbitrary reference velocity. 

\subsection{{ TaylorT1}}
\label{sec:T1}
The { TaylorT1} approximant refers to the choice corresponding 
to leaving the PN expansions of the luminosity
${\cal F}(v)$ and $E'(v)$ as they appear in Eq.~(\ref{eq:phasing formula general2})
as a ratio of polynomials and solving the differential equations numerically
\begin{subequations}
\begin{eqnarray}
\label{eq:phasing formula1a}
\frac{d\phi^{{\rm(T1)}}}{dt} - \frac{v^3}{M} & = & 0, \\
\frac{dv}{dt}  + \frac{{\cal F}(v)}{ME'(v)} & = & 0.
\label{eq:phasing formula1b}
\end{eqnarray}
\label {eq:phasing formula1}
\end{subequations}
In the above $v\equiv v^{({\rm T1})}$ but for the sake of  notational 
simplicity
we write only $v;$ from the context the meaning should be clear.
In the formulas of this section, and in the sections that follow,
the expressions for  ${\cal F}(v)$ [$E(v)$] are to be truncated
at relative PN orders 2[2], 3[3] and 3.5[3] to obtain 2PN
~\cite{Blanchet1995,Blanchet:1995fg,Will:1996zj,DIS01} , 3PN and 3.5PN
~\cite{Blanchet:2004bb,Blanchet:2004ek,Damour:2002kr}
template or signal models respectively.

To see how to set up initial conditions, refer to 
Eq.\ (\ref{eq:phasing formula general1}). 
Let the initial  gravitational wave frequency be $F_0$ or,
equivalently,  initial velocity $v_0=(\pi M F_0)^{1/3}.$ One normally
chooses $t=0$ at $v=v_0.$ This can be achieved by choosing $v_{\rm ref} 
= v_0$ and $t_{\rm ref}=0,$ in Eq.\ (\ref{eq:phasing formula general1}). 
The initial phase $\phi_{\rm ref}$ is chosen to be either $0$ or $\pi/2$
in order to construct two orthogonal templates 
(see Sec.\ \ref{sec:maximization} for details). 

\subsection{{ TaylorT4}}
\label{sec:T4}
TaylorT4 was proposed in Ref.~\cite{BCV2} and 
investigated in Refs.~\cite{Buonanno-Cook-Pretorius:2007,Baker:2006ha,Boyle:2007ft}, thus many years after 
the other approximants discussed in this paper were proposed (with the exception of TaylorEt, 
which is even more recent). However, it is a straightforward extension of TaylorT1 
and at 3.5PN order by coincidence is found to be 
in better agreement with numerical simulations of the inspiral 
phase~\cite{Buonanno-Cook-Pretorius:2007,Pan:2007nw,Baker:2006ha,Boyle:2007ft,Damour:2007yf,Gopakumar:2007vh,Damour:2008te}.  
The approximant is obtained by expanding the 
ratio of the polynomials ${\cal F}(v)/E'(v)$  to the consistent 
PN order. The equation for $v^{({\rm T4})}(t)\equiv v(t)$ at 3.5PN order 
reads,
\begin{widetext}
\begin{eqnarray}
\frac{dv}{dt}&=&\frac{32}{5}  \frac{\nu}{M} v^9 \left[1 - \left(
\frac{743}{336} + \frac{11  }{4}\nu \right) v^2 +4 \pi v^3 +\left(
\frac{34103}{18144} + \frac{13661  }{2016}\nu +\frac{59 }{18}\nu^2 \right) v^4 
   - \left( \frac{4159}{672} +\frac{189 }{8}\nu \right)\pi v^5 \right.
\nonumber\\ 
   & + & \left( \frac{16447322263}{139708800} 
+\frac{16 }{3}\pi^2 - \frac{1712 }{105}\gamma
+\left(\frac{451   }{48} \pi^2 - \frac{56198689 }{217728}\right)\nu
   + \frac{541 }{896}\nu^2 -\frac{5605 }{2592}\nu^3
-\frac{856}{105} \log (16v^2)
\right) v^6 \nonumber\\
   & - & \left. \left(  \frac{4415}{4032}
-\frac{358675 }{6048}\nu
-\frac{91495   }{1512}\nu^2
\right)\pi v^7 \right]\,.
\label{eq:dvdt t4}
\end{eqnarray}
\end{widetext}
The orbital phase $\phi^{({\rm T4})}$ is determined, as in the case
of TaylorT1, by Eq.\ (\ref{eq:phasing formula general2a}) and
 numerical solution of Eq.\ (\ref{eq:dvdt t4}) and 
(\ref{eq:phasing formula general2a}) yields the TaylorT4 approximant.
 
Note that although TaylorT1 and TaylorT4 are perturbatively 
equivalent, the evolution of the phase can be quite different 
in these two approximations. The asymptotic structure of the
approximants are also quite different: while $\dot v$ can
have a pole (although not necessarily in the region of interest)
when using Eq.\ (\ref{eq:phasing formula1b}) none is possible 
when Eq.\ (\ref{eq:dvdt t4}) is used. Differences of this
kind can, in principle, mean that the various PN families give
different phasing of the orbit. The hope is that when the PN
order up to which the approximation is known is large, then the
difference between the various PN families becomes negligible.

Setting up the initial conditions for TaylorT4 
is the same as in the case of TaylorT1. 

\subsection{{ TaylorT2}}
\label{sec:T2}
TaylorT2 is based on the second form
of the phasing relations Eq.~(\ref{eq:phasing formula general1}).
Expanding the ratio of the polynomials ${\cal F}(v)/E'(v)$ in these
equations to  consistent PN order and integrating them one obtains
a pair of parametric equations for $\phi(v)$ and $t(v)$, 
the { TaylorT2} model.
\begin{subequations}
\begin {eqnarray}
\phi^{({\rm T2})}_{n/2}(v) & = & \phi^{({\rm T2})}_{\rm ref} +
\phi^v_N (v)\sum_{k=0}^{n} \hat{\phi}^v_k v^k, \\
t^{({\rm T2})}_{n/2}(v) & = & t^{({\rm T2})}_{\rm ref} +t^v_N(v) 
\sum_{k=0}^{n} \hat{t}^v_k v^k.
\end {eqnarray}
\label{eq:phasing formula2}
\end{subequations}
Of all models considered in this study, TaylorT2 is computationally
the most expensive. This is because the phase evolution involves
solving a pair of transcendental equations which is very time-consuming. 
\begin{widetext}
\begin{subequations}
\begin {eqnarray}
\label{eq:phiref t2}
\phi^{({\rm T2})}_{3.5}(v) &=&  \phi^{(2)}_{\rm ref} 
      - \frac{1}{32\nu v^5}\left[1+ \left(\frac{3715}{1008} 
      + \frac{55}{12}\nu\right)v^2 - 10 \pi v^3 
      + \left(\frac{15293365}{1016064} + \frac{27145}{1008 } \nu
      + \frac{3085}{144}\nu^2\right)v^4 \right . \nonumber\\
      &+&  \left (\frac{38645}{672} - \frac{65}{8 }\nu \right ) 
        \ln \left ( \frac{v}{v_{\rm lso}} \right ) \pi v^5  
      + \left\{ \frac { 12348611926451}{18776862720}
      - \frac {160}{3} \pi^2 
-\frac{1712}{21}\gamma
+ \left ( \frac {2255}{48} \pi^2 
- \frac {15737765635}{12192768} \right ) \nu \right.\nonumber\\ 
      &+& \left.\left.  \frac {76055}{6912} \nu^2 
      - \frac {127825}{5184} \nu^3 - \frac{856}{21}\log (16v^2) \right\} v^6 
      + \left ( \frac {77096675}{2032128} + \frac {378515}{12096}\nu
      - \frac {74045}{6048}\nu^2 \right ) \pi v^7 \right]\,,\nonumber\\ \\
t^{({\rm T2})}_{3.5}(v) &=& t^{({\rm T2})}_{\rm ref} 
      -\frac{5M}{256 \nu v^8}\left[1+ \left(\frac{743}{252} 
      + \frac{11}{3}\nu\right)v^2 - \frac{32}{5} \pi v^3 
      + \left(\frac{3058673}{508032} + \frac{5429}{504} \nu
      + \frac{617}{72}\nu^2\right)v^4 \right.\nonumber\\
      &-& \left(\frac{7729}{252} - \frac{13}{3}\nu\right)\pi v^5
      + \left\{- \frac {10052469856691}{23471078400} + \frac {128}{3} \pi^2 
      + \frac {6848}{105} \gamma 
      + \left( \frac {3147553127}{3048192} - \frac {451}{12} \pi^2 \right)\nu 
\right. \nonumber\\ 
      &-& \left.\left.\frac {15211}{1728} \nu^2 + \frac {25565}{1296} \nu^3 
      + \frac{3424}{105}\log(16v^2) \right\} v^6 
      + \left(-\frac {15419335}{127008} - \frac {75703}{756}\nu
      + \frac {14809}{378}  \nu^2\right)\pi v^7 \right]\,.
\label{eq:tref t2}
\end {eqnarray}
\label{eq:t2}
\end{subequations}
\end{widetext}

In this case, $t_{\rm ref}$ has to be chosen so that $t=0$ when $F=F_0$
or $v=v_0.$ This can be achieved most simply by solving for $t_{\rm ref},$
using Eq.\ (\ref{eq:tref t2}), substituting $v=v_0$ on the right hand side 
and putting the left side to zero.

\subsection{{ TaylorT3}}
\label{sec:T3}
This form of the approximant goes a step further than the previous
{ TaylorT2} approximant. 
After computing as before a parametric representation
of the phasing formula $\phi(v)$ and $t(v)$, one explicitly inverts
$t(v)$ to obtain $v(t)$ and uses it to produce an explicit representation
of $\phi(t)\equiv\phi(v(t)))$. This is the { TaylorT3} approximant:
\begin{subequations}
\begin{eqnarray}
\phi^{({\rm T3})}_{n/2}(t) & = & \phi^{({\rm T3})}_{\rm ref}+\phi_N^t \sum_{k=0}^{n}
\hat{\phi}^t_k\theta^k,\\
F^{({\rm T3})}_{n/2}(t) & = & F_N^t \sum_{k=0}^{n} \hat{F}^t_k \theta^k,
\end{eqnarray}
\label{eq:phasing formula3}
\end{subequations}
where $\theta=[\nu (t_{\rm ref}-t)/(5M)]^{-1/8}$ and
$F \equiv (2\, d \phi/dt) (2 \pi)^{-1}  =v^3/(\pi M)$ 
is the instantaneous gravitational-wave frequency.
\begin{widetext}
\begin{subequations}
\begin{eqnarray}
\label{eq:phi t3}
\phi^{({\rm T3})}_{3.5}(t) &=& \phi^{({\rm T3})}_{\rm ref}
-\frac{1}{\nu \theta^5}\left[1+ \left( \frac{3715}{8064}
+\frac{55}{96}\nu
\right)\theta^2 -\frac{3\pi}{4}\theta^3
+ \left(\frac{9275495}{14450688}+\frac{284875}{258048 }\nu +
\frac{1855}{2048 }\nu^2\right)\theta^4 \right.\nonumber\\
&+& \left (\frac {38645}{21504} - \frac{65}{256 }\nu \right ) 
\ln \left ( \frac {\theta}{\theta_{\rm lso}} \right ) \pi\theta^5 
+ \left\{ \frac {831032450749357}{57682522275840} - \frac {53}{40}\pi^2
+ \left (- \frac {126510089885}{4161798144}
+ \frac {2255}{2048} \pi^2 \right ) \nu \right. \nonumber \\ 
&-& \frac {107}{56} \gamma 
+ \left.\left. \frac {154565}{1835008} \nu^2 - \frac {1179625}{1769472} \nu^3
- \frac {107}{56}\log(2\theta) \right \} \theta^6 
+ \left ( \frac {188516689}{173408256} + \frac {488825}{516096} \nu
- \frac {141769}{516096} \nu^2 \right )\pi\theta^7 \right]\,,\nonumber\\ \\
F^{({\rm T3})}_{3.5}(t) & = &  
      \frac{\theta^3}{8\pi M}\left[1+ \left(\frac{743}{2688}
+\frac{11}{32}\nu\right)\theta^2 -\frac{3}{10}\pi\theta^3 
+ \left(\frac {1855099}{14450688} + \frac{56975}{258048 }\nu +
\frac{371}{2048 }\nu^2\right)\theta^4 - \left(\frac{7729}{21504} 
- \frac{13}{256}\nu\right)\pi\theta^5 \right.\nonumber\\
&+&\left\{ - \frac {720817631400877}{288412611379200} + \frac {53}{200}\pi^2
+ \frac {107}{280} \gamma + \left ( \frac {25302017977}{4161798144} 
- \frac{451}{2048} \pi^2 \right ) \nu \right.\nonumber\\ 
&- & \left.\left.\frac {30913}{1835008} \nu^2 + \frac {235925}{1769472} \nu^3
+\frac {107}{280} \log(2\theta) \right \} \theta^6
+ \left (- \frac {188516689}{433520640} - \frac{97765}{258048} \nu
+ \frac {141769}{1290240} \nu^2 \right ) \pi\theta^7 \right]\,.
\label{eq:freq t3}
\end{eqnarray}
\label{eq:t3}
\end{subequations}
\end{widetext}

The initial conditions in this case is slightly more complicated
than the previous cases.  Given an initial frequency $F_0,$ 
one numerically solves Eq.\ (\ref{eq:freq t3}) to find the value of 
$t_{\rm ref}$ at which $F=F_0$ and $t=0$ (recall that $\theta$ 
involves $t_{\rm ref}.$) Note that as $t\rightarrow t_{\rm ref},$
formally $F\rightarrow$ diverges.

\subsection{{ TaylorEt}}
\label{sec:TEt}
The { TaylorEt} was recently introduced in 
Ref.~\cite{Gopakumar:2007jz,Gopakumar:2007vh,Tessmer:2008tq}.
Introducing\footnote{Note that the $\zeta$ in this paper is denoted variously
by $\zeta$ in~\cite{Gopakumar:2007jz} but  by $\xi$
in  e.g.~\cite{Bose:2008ix}.} $\zeta=-2 E/\nu$ (recall that our $E$ is conserved energy
per total mass), the { TaylorEt} approximants are obtained
starting from Eq.~(\ref{eq:energy}) for $E(x)$ or $\zeta(x)$ and inverting it to
obtain $x(\zeta)$:
\begin{widetext}
\begin{equation}
x=\zeta\left[1+\left(\frac{3}{4}+\frac{1}{12}\nu\right)\zeta
+\left(\frac{9}{2}-\frac{17}{8}\nu+\frac{1}{18}\nu^2 \right)\zeta^2+
\left(\frac{405}{16}+\left( \frac{205}{96}\pi^2 -\frac{4795}{72} \right)
\nu
+\frac{55}{64}\nu^2+\frac{35}{1296}\nu^3 \right) \zeta^3\right].
\label{eq:x Et}
\end{equation}
With this choice of variable the equation determining the evolution
of $v$, Eq.\ (\ref{eq:phasing formula general2b}), 
 transforms to the balance equation for $E$ rewritten in terms
of the $\zeta$ variable:
\begin{equation}
\frac{d\zeta}{dt}=\frac{{2\,\cal{F}}(v(\zeta))}{\nu\,M}.
\end{equation}
\end{widetext}
There is no difference between T1 and T4  approximants in
the Et-parametrisation
and the  gravitational-wave phasing equations
Eq.\ (\ref{eq:phasing formula general2a}) and 
Eq.\ (\ref{eq:phasing formula general2b}) in terms of $\zeta$ become 
~\cite{Bose:2008ix}, 
\begin{widetext}
\begin{subequations}
\label{Eq:Phas_Et_dphidt}
\begin{eqnarray}
\label{eq:phi Et}
\frac{d \phi^{({\rm Et})} (t)}{dt} & = & \frac{ \zeta^{3/2}}{M} 
\left [1+ \left( \frac{9}{8}+\frac{1}{8}\nu \right) \zeta
+ \left( {\frac {891}{128}} -{\frac {201}{ 64}}\,\nu
+ \frac {11}{128}\,{\nu}^{2} \right) \zeta^2 
+ \left \{ \frac {41445}{1024} - \left  ( \frac {309715}{3072}
- \frac {205}{64}\,{\pi}^{2} \right ) \nu \right . \right. \nonumber\\ 
& + & \left. \left. {\frac {1215}{1024}}\,{\nu}^{2}
+ {\frac {45}{1024}}\,{\nu}^{3} \right \} {\zeta}^{3} \right ],\\
\frac{d\zeta}{dt} & = &\frac{64 \nu  \zeta ^5 }{5 M} 
\left[ 1 + \left(\frac{13}{336}-\frac{5  }{2}\nu\right) \zeta 
+ 4 \pi  \zeta ^{3/2} + \left(
 \frac{117857}{18144} -\frac{12017  }{2016}\nu +\frac{5 }{2}\nu^2 \right)
 \zeta ^2 +\left(\frac{4913}{672} 
- \frac{177 }{8}\nu\right)\pi \zeta^{5/2} \right. \nonumber \\ 
&+& \left(
\frac{37999588601}{279417600} 
+\frac{16 }{3}\pi^2 - \frac{1712 }{105}\gamma
+\left( \frac{369   }{32}\pi^2-\frac{24861497 }{72576} \right)\nu
+\frac{488849 }{16128}\nu^2 -\frac{85 }{64}\nu^3
- \frac{856}{105} \log (16 \zeta )
\right)\zeta^3 
  \nonumber \\
&+&\left. \left( \frac{129817}{2304}
-\frac{3207739}{48384}\nu +\frac{613373   }{12096}\nu^2 \right)\pi \zeta ^{7/2} \right].
\label{eq:xi Et}
\end{eqnarray}
\label{eq:Et}
\end{subequations}
\end{widetext}
To set up the initial condition note that $2\pi F =2\, d\phi/dt.$
Given an initial frequency $F_0$ one finds the initial value 
$\zeta_0$ of $\zeta$ by numerically solving Eq.\ (\ref{eq:phi Et}),
by setting the left hand side to $\pi F_0.$

\subsection{{ TaylorF2}}
\label{sec:TF2}
The most commonly used form of the approximant is the Fourier
representation computed using the {\it stationary phase approximation} (SPA).
Using the SPA the waveform in the frequency domain may be written as,
\begin {equation}
\tilde{h}^{\rm spa}(f)= \frac {a(t_f)} {\sqrt {\dot{F}(t_f)}}
e^{ i\left[ \psi_f(t_f) -\pi/4\right]},\ \
\psi_f(t) \equiv  2 \pi f t - 2\,\phi(t),
\label{eq:ft phase}
\end {equation}
where $t_f$ is the saddle point defined by solving for $t$,
 $ d \psi_f(t)/d t = 0$,
i.e. the time $t_f$ when the gravitational-wave  frequency $F(t)$ becomes equal to the
Fourier variable $f$. In the adiabatic approximation,
(denoting $v_f \equiv (\pi M f)^{1/3}$)
the value of $t_f$ and $ \psi_f(t_f)$ are given by the following integrals:
\begin{subequations}
\begin{eqnarray}
t_f &=& t_{\rm ref} + M \int_{v_f}^{v_{\rm ref}} \frac{E'(v)}{{\cal {\cal F}}(v)}
dv,\\
 \psi_f(t_f) &=& 2 \pi f t_{\rm ref} - \phi_{\rm ref} +2  \int_{v_f}^{v_{\rm ref}}
(v_f^3 - v^3)
\frac{E'(v)}{{\cal {\cal F}}(v)} dv.\nonumber\\
\end{eqnarray}
\label{eq:s1}
\end{subequations}
As in the time domain case it is more efficient to use the equivalent
differential form
\begin{equation}
\frac{d\psi}{df} - 2\pi t = 0, \ \ \ \
\frac{dt}{df} + \frac{\pi M^2}{3v^2} \frac{E'(f)}{{\cal F}(f)} = 0,
\label {eq:frequency-domain ode}
\end{equation}
and this characterizes the { TaylorF1} approximant.

The analogue of the { TaylorT2} in the frequency domain follows by
explicitly truncating the energy and flux functions to consistent post-Newtonian orders and explicating the $v$- integration in the above.
This leads us to a Fourier domain waveform, the { TaylorF2},
 which is the most often employed  PN-approximant, given by
\begin{equation}
\tilde{h}(f) = {\cal A} f^{-7/6} e^{i \psi(f)},
\label{eq:RWF}
\end{equation}
where ${\cal A} \propto {\cal M}^{5/6} Q(\mbox{angles})/D$, and
$D$ the distance to the binary.
To 3.5PN order the phase of the Fourier domain waveform is given by
\begin{widetext}
\begin{eqnarray}
\psi^{({\rm F2})}_{3.5}(f) & = & 2\pi f t_c-\phi_c-\frac{\pi}{4}
+\frac{3}{128\,\nu\, v^5}\; \left[ 1 +\frac{20}{9}
\left( \frac{743}{336} + \frac{11}{4}\nu \right)v^2 
- 16\pi v^3 + 10\,\left( \frac{3058673}{1016064} 
+ \frac{5429\, }{1008}\,\nu + \frac{617}{144}\,\nu^2 \right)v^4
\right.\nonumber\\ 
&+& \pi\left(\frac{38645 }{756} - \frac{65}{9}\nu\right)
\left\{1  + 3\log \left(\frac{v}{v_{\rm lso}}\right)\right\} v^5
+ \left\{ \frac{11583231236531}{4694215680} - \frac{640}{3}\pi^2 -
\frac{6848\,\gamma }{21} -\frac{6848}{ 21} \log\left(4\;{v}\right) 
\right. \nonumber\\ 
&+& \left.\left.\left( - \frac{15737765635}{3048192} 
+ \frac{2255\,{\pi }^2}{12} \right)\nu 
+\frac{76055}{ 1728}\nu^2-\frac{127825}{ 1296}\nu^3
\right \} v^6 
+ \pi\left(\frac{77096675 }{254016} + \frac{378515}{1512}\,\nu 
- \frac{74045}{756}\,\nu^2\right)v^7 \right],\nonumber\\ 
\label{eq:3.5PN-phasing}
\end{eqnarray}
\end{widetext}
where $v= (\pi M f)^{1/3}$.  

In this case one has to specify the constants $t_c$ and $\phi_c$ 
and they can be chosen arbitrarily. 

\subsection{The effective-one-body model}
\label{sec:EOB}

In this paper since we are not particularly concerned with the coalescence signal, we employ the less sophisticated 
earlier version of the EOB model calibrated to numerical-relativity simulations from 
Ref.~\cite{Buonanno:2007pf} (for more sophisticated versions of the EOB model  
see Refs.~\cite{Damour:2007vq,Damour:2008te,Boyle2008a,Damour:2009b,Buonanno:2009qa}). 
Below we briefly review the EOB model from Ref.~\cite{Buonanno:2007pf}.

Introducing polar coordinates $(r,\phi)$ and their conjugate momenta $( p_r,p_\phi)$, the EOB effective metric takes the 
form~\cite{BuonD98}
\begin{equation}
  ds_{\rm eff}^2 =
  -A(r)\,dt^2 + \frac{D(r)}{A(r)}\,dr^2 +
  r^2\,\Big(d\theta^2+\sin^2\theta\,d\phi^2\Big) \,.  
\label{eq:EOBmetric}
\end{equation}
The EOB Hamiltonian reads
\begin{equation}
\label{himpr}
H^{\rm real}(r,p_{r},p_\phi) \equiv \mu\hat{H}^{\rm real} 
= M\,\sqrt{1 + 2\nu\,\left ( \frac{H^{\rm eff} - \mu}{\mu}\right )} \,,
\end{equation}
with the effective Hamiltonian~\cite{BuonD98,Damour:2000we}
\begin{equation}
  \label{eq:genexp}
  H^{\rm eff}(r,p_{r},p_\phi) \equiv \mu\,\widehat{H}^{\rm eff} 
  = \mu\,\sqrt{A (r) \left[ 1 + \frac{A(r)}{D(r)} p_r^2 + \frac{p_\phi^2}{r^2} 
      + 2(4-3\nu)\,\nu\,\frac{p_{r}^4}{r^2} \right]} \,.
\end{equation}
The Taylor-approximants to the coefficients $A(r)$ and $D(r)$ can be
written as~\cite{BuonD98,Damour:2000we}
\begin{subequations}
\begin{eqnarray}
  \label{coeffA}
  A_{k}(r) &=& \sum_{i=0}^{k+1} \frac{a_i(\nu)}{r^i}\,,\\
  \label{coeffD}
  D_{k}(r) &=& \sum_{i=0}^k \frac{d_i(\nu)}{r^i}\,.
\end{eqnarray}
\end{subequations}
The functions $A(r)$, $D(r)$, $A_k(r)$ and $D_k(r)$ all depend on 
the symmetric mass ratio $\nu$ through the $\nu$--dependent coefficients 
$a_i(\nu)$ and $d_i(\nu)$.  These coefficients are currently
known through 3PN order (i.e. up to $k=4$) and can be read 
from Ref.~\cite{Buonanno:2007pf}. During the last stages of inspiral
and plunge\footnote{To deal with the steep rise of various quantities during
the plunge, it is advantageous to consider the EOB equations in terms of
the tortoise radial coordinate $r_*$ and its conjugate $p_{r_*}$ rather than
in terms of the standard radial coordinate $r$ and  $p_r$ as above.
The form of $H^{\rm eff}$ in the two cases will be 
different~\cite{Damour:2007xr}.
For the level of accuracy in our present work, this difference is
irrelevant.}, the EOB dynamics can be adjusted closer to the numerical
simulations by including in the radial potential $A(r)$ a p4PN 
coefficient $a_5(\nu)$ and $a_5(\nu)=\lambda_0\,\nu$, with
  $\lambda_0$ a constant\footnote{Note that $\lambda_0$ 
    was denoted $\lambda$ in Ref.~\cite{Buonanno:2007pf}, and $a_5$ in 
Refs.~\cite{Damour:2007xr,Damour:2007yf,Damour:2008te,Boyle2008a}.}. 
In order to assure the presence of a horizon in the effective metric (\ref{eq:EOBmetric}), a zero needs to
be factored out from $A(r)$. This is obtained by applying a Pad\'e 
resummation~\cite{Damour:2000we}. The Pad\'e coefficients for the expansion of
$A(r)$ and $D(r)$ at p4PN order are denoted  
$A_4^1(r)$ and $D_4^0(r)$, and their explicit form can
be read from Ref.~\cite{Buonanno:2007pf}.

The EOB Hamilton equations are written in terms of the reduced (i.e., dimensionless) quantities
  $\widehat{H}^{\rm real}$ [defined in Eq.~(\ref{himpr})], $\widehat{t} = t/M$,
and  $\widehat{\omega} = \omega\,M$~\cite{Buonanno:2000ef}: 
\begin{subequations}
  \label{eq:eobhami}
\begin{eqnarray}
  \frac{dr}{d \widehat{t}} &=& \frac{\partial \widehat{H}^{\rm real}}
  {\partial p_{r}}(r,p_{r},p_\phi)\,, 
  \label{eq:eobhamone} \\
  \frac{d \phi}{d \widehat{t}}  &=& 
  \frac{\partial \widehat{H}^{\rm real}}
  {\partial p_\phi}(r,p_{r},p_\phi)\,, \\
  \frac{d p_{r}}{d \widehat{t}} &=& - \frac{\partial \widehat{H}^{\rm real}}
  {\partial r}(r,p_{r},p_\phi) \,, \\
  \frac{d p_\phi}{d \widehat{t}} &=&
  \widehat{\cal F}_\phi(r,p_{r},p_{\phi})\,,
  \label{eq:eobhamfour}
\end{eqnarray}
\end{subequations}
with the definition
$\widehat{\omega}\equiv d \phi/d \widehat{t}$.  Another critical input to the EOB model is the form for
the radiation reaction force arising from the basic PN expression of the energy flux. 
Different choices include Pad\'e resummation~\cite{Damour1998}, and the more recent $\rho_{\ell m}$- 
resummation~\cite{Damour:2009a}. It also further includes the introduction of terms
describing next-to-quasi-circular effects. Here, for the $\phi$ component of the radiation-reaction
force we use the less sophisticated Keplerian Pad\'e-approximant to the energy
flux as given by Eq.~(15) of Ref.~\cite{Buonanno:2007pf}. 

The inspiral-plunge EOB waveform at leading order in a PN expansion reads
\begin{equation}\label{inspwave}
{h}^{\rm insp-plunge}(t) \equiv \widehat{\omega}^{1/3}\,\cos [2 \phi(t)]\,.
\end{equation}
The merger-ringdown waveform in the EOB approach is built as a superposition 
of quasi-normal modes~\cite{Buonanno:2000ef,Damour06,
Buonanno-Cook-Pretorius:2007,Buonanno:2007pf,Damour:2007xr,Damour:2007yf}, as
\begin{equation}
\label{RD}
h^{\rm merger-RD}(t) = \sum_{n=0}^{N-1} A_n\,
e^{-i\sigma_n (t-t_{\rm match})},
\end{equation}
where $n$ is the overtone number of the Kerr
quasi-normal mode, $N$ is the number of overtones included in
our model, and $A_n$ are complex amplitudes to be determined
by a matching procedure described below. 
The quantity $\sigma_{n} = \omega_{n} - i
\alpha_{n}$, where the oscillation frequencies $\omega_{n}>0$ and the 
inverse decay-times $\alpha_{n}>0$, are numbers associated with each quasi-normal mode. The
complex frequencies are known functions of the final black-hole mass
and spin and can be found in Ref.~\cite{BCW05}. The final
black-hole masses and spins are obtained from the fitting 
to numerical results worked out in Ref.~\cite{Buonanno:2007pf}.

The complex amplitudes $A_{n}$ in Eq.~(\ref{RD}) are determined
by matching the EOB merger-ringdown waveform with the EOB
inspiral-plunge waveform close to the EOB light ring. In particular, 
in Ref.~\cite{Buonanno:2007pf} the matching point is provided analytically 
by Eq.~(37). In order to do this, $N$ independent complex
equations are needed. The $N$ equations are obtained at the matching time by
imposing continuity of the waveform and its time derivatives,
\begin{equation}
\frac{d^k}{dt^k}h^{\rm insp-plunge}(t_{\rm match})=
\frac{d^k}{dt^k}h^{\rm merger-RD}(t_{\rm match})\,,\quad  
(k=0,1,2,\cdots, N-1)\,.
\end{equation}
In this paper we use N=3. The above matching approach is  referred to
as {\it point matching}. It gives better smoothness around the
matching time, but it is not very stable numerically when $N$ is large
and higher order numerical derivatives are needed. More sophisticated matching procedures 
have been proposed in the literature to overcome the stability issue. 
Reference~\cite{Damour:2007xr} introduced the 
{\it comb matching} approach where $N$ equations are obtained at $N$ points
evenly sampled in a small time interval $\Delta t_{\rm match}$ 
centered at $t_{\rm match}$. More recently, to improve the smoothness of the comb matching 
Ref.~\cite{Buonanno:2009qa} introduced the {\it hybrid comb matching} where one chooses a time interval 
$\Delta t_{\rm match}$ ending at $t_{\rm match}$, and imposes not only
 the continuity of 
the waveform at $N-4$ points evenly sampled from $t_{\rm match}-\Delta
  t_{\rm match}$ to $t_{\rm match}$, but also 
  requires continuity of the first and second order time derivatives of
  the waveform at $t_{\rm match}-\Delta t_{\rm match}$ and 
$t_{\rm match}$.

Finally, the full (inspiral-plunge-merger-ringdown) EOB waveform reads
\begin{equation}
\label{eobfullwave}
h(t) = h^{\rm insp-plunge}(t)\,
\theta(t_{\rm match} - t) + 
h^{\rm merger-RD}\,\theta(t-t_{\rm match})\,,
\end{equation}
where we denote with $\theta$ the Heaviside step function. 

\subsection{Waveforms and termination conditions}

Before concluding this Section we note a few other points concerning
the generation of the waveform. Since our goal is to study the agreement
between different waveforms it is not necessary to separately consider
the two different polarizations but only the detector response. 
For time-domain models TaylorT1, TaylorT2, TaylorT3, TaylorT4 and EOB
the waveform is taken as:
$$h_{\rm A}(t) =  C\, v_{\rm A}^2\, \sin [2\,\phi_{\rm A}(t)],$$
where $v_{\rm A}$ and $\phi_{\rm A}(t)$ are  computed using the relevant formulas corresponding to the approximant A. 
In the case of TaylorEt the waveform is taken to be
$$h_{\rm Et}(t) =  C\, \zeta (t)\, \sin [2\,\phi_{\rm Et}(t)].$$
In all cases the constant $C$ is fixed by demanding that the norm of 
the signal be unity (cf. Sec.\ \ref{sec:overlaps}).
The initial phase of the signal is set to 0, while in the case of 
templates we construct two orthonormal waveforms corresponding to 
the starting phases of $0$ and $\pi/2.$ 

\begin{table}
\caption{Termination condition for waveform generation is chosen
to be either LSO corresponding to Schwarzschild metric $v_S=6^{-1/2},$
or the extremum defined by the P-approximant of the energy function
as in~\cite{Damour1998} which is $v_{P_4}$ at 2PN and $v_{P_6}$ at 3-
and 3.5PN. 
In the case of TaylorT3 at 3.5PN, 
as the frequency evolution is not monotonic,
the evolution has to be terminated prematurely at $v_m$ such that 
$\dot{F}(v_m)=0$.}
\label{tab:ending freq}
\begin{tabular}{ccccccc}
\hline
\hline
Order/Approx & T1     & T2   & T3   & T4   & Et   & F2  \\
\hline
2PN          & $v_S$  & $v_S$& $v_S$& $v_{P_4}$& $v_{P_4}$& $v_{P_4}$   \\
\hline
3PN          & $v_S$  & $v_S$& $v_S$& $v_{P_6}$& $v_{P_6}$& $v_{P_6}$     \\
\hline
3.5PN        & $v_S$  & $v_S$& $v_m$& $v_{P_6}$& $v_{P_6}$& $v_{P_6}$     \\
\hline
\hline
\end{tabular}
\end{table}

The waveforms are terminated when $v$ reaches the value quoted 
in Table I or before, if the frequency evolution is not monotonic 
(see next Section).  For instance, in the case of TaylorT3 at 
3.5PN order the approximant has an unusual behaviour whereby 
the frequency evolution ceases to be monotonic well before 
$v$ reaches the nominal value of $1/\sqrt{6}.$ In the
case of TaylorT1, TaylorT2 and TaylorT3, the termination is at 
the LSO defined by the Schwarzschild metric, 
namely $v=1/\sqrt{6},$ at all PN orders, but we also check for
monotonicity of the frequency evolution.  For other approximants, 
except EOB, we terminate at the extremum of the P-approximant energy 
function~\cite{Damour1998}. In the case of EOB, the waveform is terminated 
at the end of the quasi-normal ringing.

\section{Frequency evolution}
\label{sec:freq evolution}
The quantity that determines the evolution of a binary, its phasing
and the duration for which it lasts starting from a particular frequency,
is the acceleration of the bodies under radiation reaction. Equivalently,
it is the evolution of the derivative of the 
gravitational wave frequency $\dot F = dF/dt,$ 
which determines the phasing of the waves.  When the separation 
between the bodies is large, the frequency evolution is slow and the 
quantity~\cite{DIS01} $\epsilon(t) = \dot F F^{-2},$ which measures the fractional change in the
 frequency over a period, is small: $\dot F F^{-2}\ll 1.$
As the binary evolves, this quantity increases but, as seen in numerical
evolutions, remains finite and positive all the way up to the merger of
the two bodies. In what follows we will explore the behaviour of $\epsilon$
as a function of the PN parameter $v$ rather than $t,$ because 
the former parameter is (mass) scale free, unlike the latter. 

Computing the adiabaticity parameter $\epsilon(v)$ in the case of
TaylorT1 and TaylorT4 is straightforward using  Eqs.\
(\ref{eq:phasing formula1b}) and (\ref{eq:dvdt t4}).
In the case of TaylorT2, one differentiates Eq.\ (\ref{eq:tref t2})
with respect to $v$ and then takes its reciprocal. Finding
$\epsilon(v)$ in the case of TaylorEt is more involved. The
frequency $F$ is given by Eq.\ (\ref{eq:phi Et}) but the right
hand side is a function of $\zeta.$ One must, therefore, combine
Eqs.\ (\ref{eq:phi Et}) and (\ref{eq:xi Et}) to find the derivative
of the frequency:
\begin{equation}
\pi \dot F = \pi \frac{dF}{d\zeta} \frac{d\zeta}{dt} = 
\frac{d}{d\zeta} \left ( \frac{d\phi}{dt}\right)\frac{d\zeta}{dt}.
\end{equation}
The above equation still gives $\dot F$ as a function of $\zeta.$
One can then use Eq.\ (\ref{eq:x Et}) to get $\epsilon(v).$
Consequently, there is no guarantee that $v$ will be monotonic 
in the region of interest. However, we do find that the function
$\epsilon_{\rm Et}(v)$ is positive in the region of
interest and therefore $v$ increases monotonically for TaylorEt. 
To find $\epsilon(v)$ for TaylorT3, $\dot F$ is given by differentiating 
Eq.\ (\ref{eq:freq t3}) with respect to $t$ (recall $\theta=\theta(t)$) 
and then one uses the same equation to find $v=(\pi M F)^{1/3}$ at a given $t.$
It turns out that for TaylorT3 the function $\epsilon_{\rm T3}$ can become
negative in the region of interest (exactly when this happens 
depends on the PN order and mass ratio) 
and so $v$ does not generally increase monotonically.

\begin{figure*}
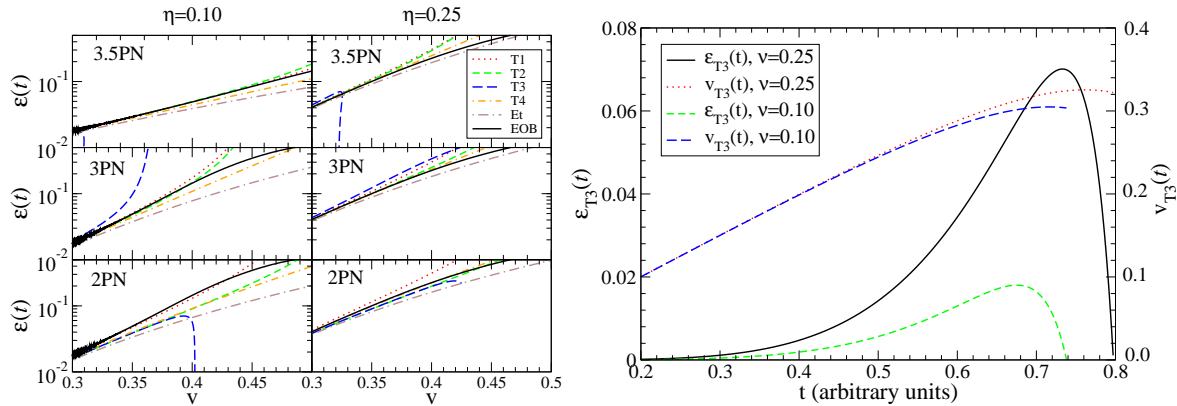

\includegraphics[width=0.41\textwidth]{fdot/fdot.eps}
\includegraphics[width=0.45\textwidth]{fdot/fdot-T3.eps}
\caption{On the left hand panel the plots show the evolution of 
frequency in different PN families. The adiabaticity 
parameter $\epsilon(t) \equiv F^{-2}\dot F$ 
is essentially the same for all the different approximations at 
$v\ll 1$ As the binary gets close to coalescence the various  
approximations begin to differ from each other.
The right hand panel shows the adiabaticity parameter as a
function of time $t$ at 3.5PN order. Note that $\epsilon_{\rm T3}(t)$ 
begins to decrease and even becomes less than zero before
$v$ reaches its nominal value of $1/\sqrt{6}.$ This leads
to waveforms that are significantly shorter in the
case of TaylorT3.}
\label{fig:freq evolution}
\end{figure*}

Figure~\ref{fig:freq evolution}, left panel, plots $\epsilon(v)$ 
for two values of the mass ratio: $\nu=0.10$ and $\nu=0.25.$ 
When $v$ is small ($v\ll 1/\sqrt{6}$)
$\epsilon(v)$ for the different approximants is the same. 
Therefore, in systems for which $v$ remains small when the signal
is in band (as, for example, in a binary neutron star), the different 
approximants, as we shall see in the next Section, agree well
with each other. As $v$ approaches $1/\sqrt{6},$ different
approximations tend to differ greatly, which means we cannot
expect good agreement between the different PN families.
Of the approximants considered here, TaylorEt  seems to have
the smallest value of $\epsilon(v)$ at any given $v.$ Therefore,
the evolution will be slower, and the duration of the waveform
from a given frequency larger, than the other approximants
~\cite{Tessmer:2008tq}. TaylorT3 also differs from all others because
$\epsilon(v)$ becomes negative before the last stable orbit, and so
$v$ does not generally increase monotonically for this approximant. 
This behavior can be seen at 2PN and 3.5PN orders in the left panel of 
Fig.~\ref{fig:freq evolution}.
The reason for this can be seen in Fig.\ \ref{fig:freq evolution}, right panel, 
where we have shown the time development of $\epsilon_{\rm T3}(t)$ for two values of $\nu
=0.10,\, 0.25.$ Since $\dot F$ becomes negative before reaching
the last stable orbit,  the waveform has to be terminated before
before $v$ reaches $1/\sqrt{6}.$ 
 
\begin{figure}
\includegraphics[width=3.8truein]{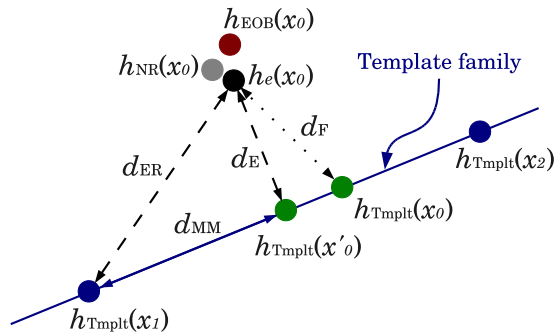}
\caption{Schematic plot of distance (or mismatch) relation between 
templates and exact, numerical and EOB waveforms. 
}
\label{fig:mismatch}
\end{figure}

\section{Effectualness}
\label{sec:overlaps}

The goal of this study is to compare the different PN approximations
by measuring their mutual effectualness (i.e., overlaps maximized
over intrinsic and extrinsic parameters) for a number of different mass pairs.
To this end it will be very useful to define the scalar product of
waveforms.  Given waveforms $h_k$ and $q_k,$ $k=0,\ldots, N-1,$
where $h_k$ is the $k$th sample of the signal $h(t)$ at time 
$t_k=k\Delta,$ $\Delta=1/f_s$ being the sampling interval corresponding 
to the sampling rate $f_s,$ their scalar product is defined by\footnote{It
is conventional to define the scalar product in the continuum limit. Here,
however, we have given the definition for a discretely sampled data and 
this is the expression that is used in computing the overlaps.}
\begin{equation}
\left < h,\, q \right > (\tau_k) = 2 \sum_{m=0}^{N-1} \left [ 
H_m\,Q^*_m + H^*_m\,Q_m \right ]\, e^{-2\pi i m k /N}\, 
\frac{\Delta_f}{S_h(f_m)}
\end{equation}
where $\Delta_f=f_s/N,$ $f_m=m\Delta_f,$ $\tau_k=k\Delta$ is the lag 
of the template --- a measure of the relative time-shift between the 
template and signal, $H_m =\Delta\sum_{k=0}^{N-1} h_k\, e^{2\pi i m k/N}$
is the discrete Fourier transform of $h(t)$ 
(similarly, $Q_m$) and $S_h(f_m)$ is the one-sided noise power 
spectral density of a detector.  In comparing two waveforms the 
overall amplitude is of no interest and we should, therefore, 
consider waveforms with unit norm, namely 
$\hat h = h/\sqrt{\left<h,\,h\right >}.$
Consequently, the relevant quantity is the scalar product between 
normalized waveforms defined by
\begin{equation}
{\cal O}[h,\, q] = \left < \hat h,\, \hat q \right >
\end{equation}

\subsection{Maximization of the overlaps}
\label{sec:maximization}

The signal and the template both depend on a set of parameters of
the source (e.g., masses and initial spins of the component masses) 
and its orientation relative to the detector. We shall be concerned 
with binaries with non-spinning components on quasi-circular orbits.
Such systems are characterized by two intrinsic parameters, namely
the masses $m_1$ and $m_2$ of the components, and two extrinsic 
parameters, namely the time-of-coalescence $t_C$ and the phase of 
the signal at that time $\phi_c.$ The overlap integral, therefore,
depends on the parameters of the signal and the template and the
relevant quantity is the overlap maximized over these parameters.

The data analysis problem is concerned with digging out a specific
signal buried in noisy data.  This means that the parameters of
the signal are fixed but the data analyst is at liberty to maximize
over the parameters of the template. In this paper we will 
explore the effectualness of templates; that is to say the overlap 
maximized over a template's parameters keeping
those of the signal fixed.  We will do this for several choices of
the component masses of the binary. However, the time-of-coalescence
$t_C$ and the phase $\phi_C$ of the signal at that time, are arbitrarily
chosen to be equal to zero. A caveat is in order concerning the value of
the effectualness arising as a result of our choice of $t_C$ and $\phi_C:$ 
the maximized overlap is not very sensitive to our choice of 
$t_C$ but it could vary by several percents depending on the choice
of a signal's phase, especially when the signal and the template 
families are not very close in the geometrical sense.

Maximization over a template's masses is carried out using a bank 
of templates and the template bank is set up such that for all 
signals of the same family as the template their best overlap with 
the nearest template is larger than a certain value called the
minimal match ${\rm MM}$.  Our template placement is as in Ref.\
\cite{Cokelaer:2007kx}, which is known to produce, with probability close
to 1 \cite{Cokelaer:2007kx}, matches larger than the minimal match for 
the TaylorT1, TaylorT3, TaylorF2 and EOB families of signals (and templates)
for the range of masses considered in this paper. 
We have checked this to be true also for TaylorEt and TaylorT4 families.

We have used a minimal match of ${\rm MM}=0.99$ in
all cases. Maximization over time-of-coalescence is accomplished
by looking at the overlap integral at different lags $\tau_k.$ Finally,
since our templates are of the form $h_k = A_k \cos(\phi_k+\phi_0),$ where
$\phi_0$ is an unknown constant phase offset, maximization over 
$\phi_0$ can be achieved by using two quadratures of the template,
$h_k^0 = A_k\cos(\phi_k)$ and $h_k^{\pi/2} = A_k\cos(\phi_k+\pi/2):$ 
\begin{equation}
\max_{\phi_0} {\cal O}[h,\,q] = \left [ \left <h^0,\,q \right >^2
+ \left <h^{\pi/2},\,q \right >^2\right ]^{1/2}. 
\end{equation}
When the signal and the template belong to the same family the 
maximized overlap is at least ${\rm MM}$. When the waveforms belong to different 
families the maximized overlap is less than ${\rm MM}$. 

Our approach to finding the effectualness of a template with a signal of
``fixed'' parameters is here somewhat different from what
is normally followed in the literature, but more appropriate in the 
context of data analysis. In the literature on the comparisons of 
different PN models, one normally measures either the {\it best}
or the {\it minimax} overlap~\cite{Damour1998}. The best overlap gives the maximum of 
the overlap over the masses and $t_C$ but maximized over the 
constant phases of both the signal and the template. On the other
hand, the minimax overlap is the overlap maximized over the
masses and $t_C$ but minimized over the constant phases of the
signal and the template.  As mentioned earlier, we fix the phase 
of the signal to be equal to zero and hence our effectualness is, in 
principle, smaller than best overlaps but larger than minimax overlaps. 
The difference between the best and minimax overlaps is tiny when 
the effectualness is intrinsically large (i.e., close to 1), but could
differ by $5-8\%$ when the best overlap is $\sim 0.8.$ This should
be kept in mind while interpreting our results.  Moreover, as 
mentioned earlier, instead of numerically searching for the 
maxima of the overlap in the space of masses we just use a 
grid of templates with a minimal match of ${\rm MM}=0.99.$ 

We will compute effectualness between every possible template 
and signal. If our template is the PN approximation $A$ and the signal
is the PN approximation $B$ then we are interested in computing the
matrix $\epsilon_{AB}$ defined by
\begin{equation}
\epsilon_{AB} \equiv \max_{\lambda^A} {\cal O}[h_A(\lambda^A),\, h_B(\lambda^B)],
\end{equation}
where $\lambda^A$ and $\lambda^B$ are the parameters of the template and the
signal, respectively.  The overlap is symmetric in its arguments $h_A$ and $h_B$ 
only if the signal and template, together with their parameters, are
interchanged. That is, ${\cal O}[h_A(\lambda^A),\, h_B(\lambda^B)] = 
{\cal O}[h_B(\lambda^B),\, h_A(\lambda^A)]$ but, in general, 
${\cal O}[h_A(\lambda^A),\, h_B(\lambda^B)] \ne {\cal O}
[h_A(\lambda^B),\, h_B(\lambda^A)].$ Therefore, the maximized 
overlap $\epsilon_{AB}$ {\it need not} be symmetric. 
The process of maximization, in which
the parameters of the ``signal'' are kept fixed and those of the ``template''
are varied, breaks down the symmetry.  The lack of symmetry arises
primarily because the signal manifolds ${\cal M}_{A,B}$ representing 
the two families are distinct; the nearest ``distance'' from a 
coordinate point $P$ on ${\cal M}_A$ to a point on ${\cal M}_B$
need not be the same as the nearest distance from $P$ on ${\cal M}_B$
to a point on ${\cal M}_A.$ 
\subsection{Effectulness, faithfulness and loss in event rates}
\label{sec:eventrate}

A direct measure of the efficiency of a template bank is the loss of
event rates due to differences between the template family and the exact
signal. The loss of event rates is determined by two factors: the 
effectualness of the templates in matching the exact waveforms 
and the minimal match of the template bank itself. In this section, we will 
quantify this relation.

In Fig.\ \ref{fig:mismatch}~\footnote{This figure is very similar 
to Fig. 3 of Ref.~\cite{Lindblom:2008cm}} we sketch a portion of the
waveform space.  The solid line represents the template family subspace. Dots
represent various waveforms: (i) $h_{\rm Tmplt}(x_1)$ and $h_{\rm Tmplt}(x_2)$
are two neighboring templates in the template bank with physical
parameters $x_1$ and $x_2$; (ii) $h_{\rm Tmplt}(x_0)$ and $h_{\rm
Tmplt}(x'_0)$ are waveforms in the same family as the templates
to be chosen as discussed below; 
(iii) $h_e(x_0)$, $h_{\rm NR}(x_0)$ and $h_{\rm EOB}(x_0)$ are exact,
numerical and EOB waveforms of the same physical parameters $x_0$,
respectively. [The EOB waveform is calibrated to the numerical simulation.] 
We choose $x'_0$ such that the overlap between $h_{\rm Tmplt}(x_1)$ and
$h_{\rm Tmplt}(x'_0)$ is the minimal match (see below) of the template bank.
We choose $x_0$ such that $h_e(x_0)$ is the exact waveform that has
larger overlap with $h_{\rm Tmplt}(x'_0)$ than with any other waveforms in
the template family. This overlap is larger than the one between $h_e(x_0)$ and $h_{\rm
Tmplt}(x_0)$ even though they have the same physical parameters, because of
the systematic difference between the family of exact waveforms and the
family of templates.

We define the distance in the waveform space between two waveforms $h$ and $q$ 
by the scalar product $\sqrt{1-{\cal O}[h,\,q]}$. 
For convenience, we define the mismatch to be the square of the distance. 
The overlap between $h_{\rm Tmplt}(x_1)$ and $h_{\rm Tmplt}(x'_0)$
is the minimal match and we denote the corresponding mismatch by  $d_{\rm
MM}=1-{\rm MM}$. Similarly, $1-d_{\rm E}$ and $1-d_{\rm F}$ are the effectualness and
faithfulness of the template family with the exact waveform $h_e(x_0)$,
respectively. The mismatch between $h_e(x_0)$ and the closest template $h_{\rm
Tmplt}(x_1)$ quantifies the reduction in signal-to-noise ratio when the template
bank is used to search for the exact waveform. We denote this
mismatch by $d_{\rm ER}$. When these mismatches are small, by
Pythagorean theorem, we have an the approximate relation $d_{\rm
ER}\simeq d_{\rm MM}+d_{\rm E}$. Assuming uniform spatial distribution
of sources, 
the reduction in event rate is $1-(1-d_{\rm ER})^3\simeq 3d_{\rm ER}$.
Therefore, if we want to satisfy the usual requirement of
$<10\%$ loss in event rate, we need $d_{\rm ER}=d_{\rm MM}+d_{\rm E}<3.5\%$.  
Typical minimal match adopted in current searches has either 
$d_{\rm MM}=3\%$ or $d_{\rm MM}=1\%$, which means,
in the first case, an extremely rigorous requirement on the
effectualness: $d_{\rm E}<0.5\%$, or in the second case, a reasonable
requirement of $d_{\rm E}<2.5\%$. The latter is achievable by PN models. 
Note that, if both the minimal match of a template bank 
and the effectualness of the template model are $97\%$, the loss in 
event rate rises to $17\%$.

However, it is not possible to calculate $d_{\rm ER}$ since we do not know 
the exact waveform $h_e(x_0)$. In this paper, we adopt two 
strategies to estimate $d_{\rm ER}$: (i) we calculate the mutual effectualness 
of PN models for low-mass binaries and assume it to be a good 
representation of their effectualness with exact
waveforms; (ii) we approximate $h_e(x_0)$ with the EOB waveform 
$h_{\rm EOB}(x_0)$ calibrated to the numerical simulations. We can verify the goodness of the latter assumption 
as follows. The mismatch between the best EOB waveforms~\cite{Buonanno:2009qa, Damour:2009b} and the numerical 
waveforms is less than $10^{-3}$. In Ref.\ \cite{Buonanno:2009qa}, the authors calculated the mismatch 
among accurate numerical waveforms generated by simulations with
different resolutions and/or extraction schemes, as well. They found that the mismatch 
is less than $10^{-4}$. We consider the latter as an estimate of the mismatch
between exact and numerical waveforms. In the worst case, the mismatch between the exact and EOB
waveforms with the same physical parameters is roughly
$(\sqrt{10^{-3}}+\sqrt{10^{-4}})^2=1.7\times 10^{-3}$. Therefore, we 
can conclude that by approximating $h_e(x_0)$ with $h_{\rm EOB}(x_0)$.
we underestimate the loss of event rate by at most $0.5\%$.

Notice that the effectualness result presented in the following sections is
slightly different from $1-d_{\rm E}$. It is obtained through discrete
searches over template parameters using template banks with ${\rm MM}=0.99$
rather than through continuous searches. Therefore, the mismatch associated
with this effectualness result includes already the discreteness effect
in the template banks, i.e. a mismatch $d_{\rm MM}^{(0)}=0.01$. In this
case, if a search
is carried out with a template bank of a different minimal match,
say MM$=1-d_{\rm MM}=0.97$, to calculate the loss of event-rate, a
mismatch of $d_{\rm MM}-d_{\rm MM}^{(0)}=0.02$, instead of $d_{\rm MM}$,
needs to be added to the effectualness result in this paper, i.e. $d_{\rm ER}=d_{\rm MM}-d_{\rm MM}^{(0)}+d_{\rm E}$. 
The only exception in this paper is the effectualness result between EOB
models presented in the Conclusions which is obtained
through a continuous search.

\subsection{Choice of binary systems and PN orders}
\label{sec:systems}

We have chosen three conventional systems, binary neutron stars (BNS),
binary black holes (BBH) and binary neutron star-black hole systems, 
but we have chosen the BNS and BBH systems to be slightly asymmetric,
$(1.38,\, 1.42)M_\odot$ and $(9.5,\, 10.5)\,M_\odot$ but NS-BH is 
chosen to be the conventional $(10,\,1.4)\,M_\odot$ system. To this
we have added another binary with component masses $(4.8,\, 5.2)$
which lies on the border line between where most PN families
are similar to one another and where they begin to differ.

We compute overlaps maximized over a template bank between seven 
different models { (TaylorT1, TaylorT2, TaylorT3,
TaylorT4, TaylorF2, TaylorEt, EOB)}, each at three different
PN orders ($v^4,\, v^6,\, v^7$). The results will be presented in the
form of a set of Figures. For each mass pair there will 
be one Figure consisting of 9 panels (one panel for each PN order),
each panel containing seven curves (one each for each template
family at that order) and each curve with 21 data points corresponding
to signals from the seven PN families at each three different PN
orders, 2PN, 3PN and 3.5PN. 

\section{Results of the effectualness of PN templates}
\label{sec:results}

\begin{figure*}[t]
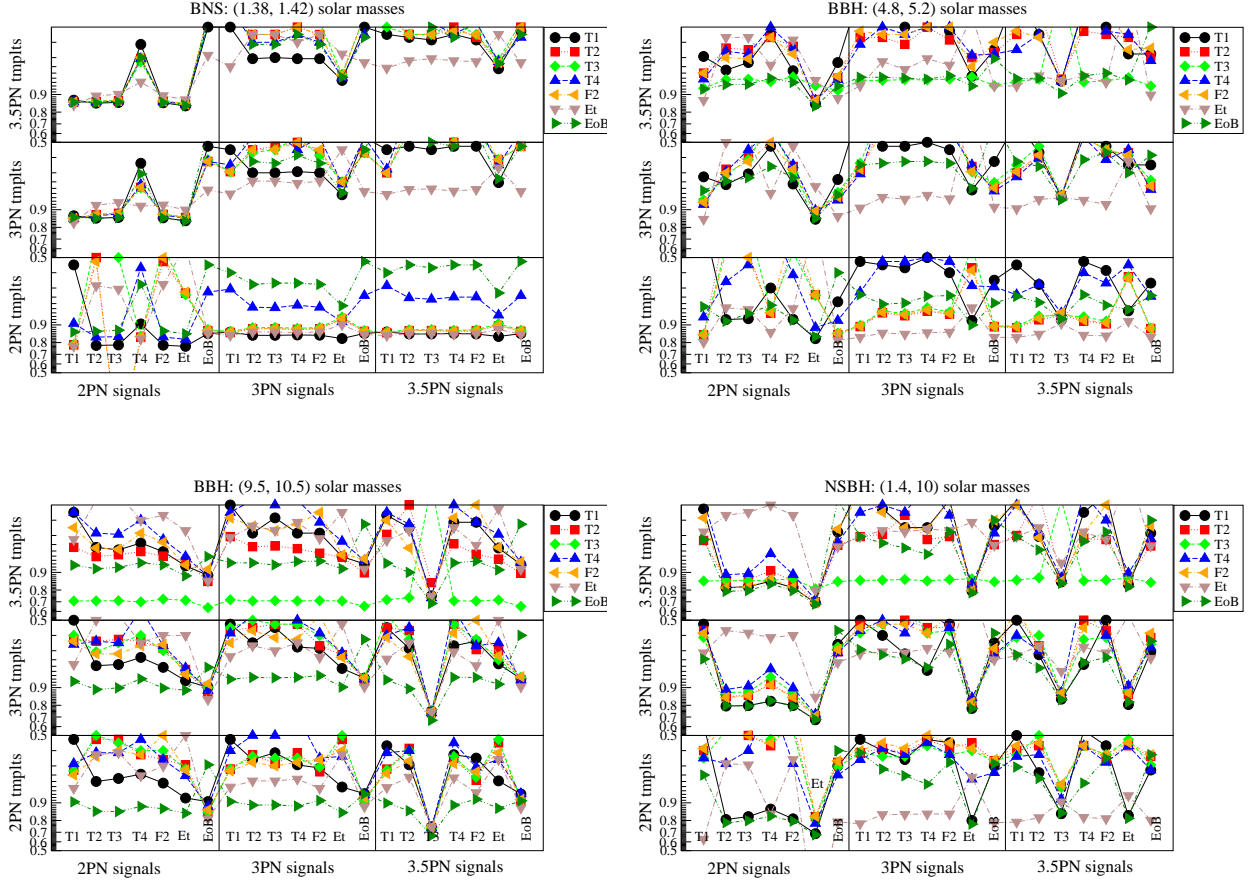

\includegraphics[width=0.45\textwidth]{overlaps/ini-ligo/ini-BNS.eps}
\hskip0.20cm
\includegraphics[width=0.45\textwidth]{overlaps/ini-ligo/ini-BBH2.eps}
\vskip1.0cm
\includegraphics[width=0.45\textwidth]{overlaps/ini-ligo/ini-BBH1.eps}
\hskip0.20cm
\includegraphics[width=0.45\textwidth]{overlaps/ini-ligo/ini-NSBH.eps}
\caption{The plot shows the effectualness of templates and 
signals of different post-Newtonian familes and orders for four
different binary systems for Initial LIGO.
For a template from a given PN approximation (indicated
by different line styles and symbols) and order (top
panel 3.5PN, middle panel 3PN and bottom panel 2PN) we 
compute the effectualness of each of the templates with 
signals from each of the seven families, TaylorT1 (T1), 
TaylorT2 (T2), TaylorT3 (T3), TaylorT4 (T4), TaylorF2 (F2),
TaylorEt (Et) and Effective-One-Body  (EOB), at 2PN, 3PN and 3.5PN orders. 
For instance, solid lines with filled circles give the 
effectualness of TaylorT1 templates at 3.5PN (top panel), 3PN (middle
panel) and 2PN(bottom panel) PN orders, with signals
that belong to different PN approximations and orders.
In clockwise order the panels from top left correspond to
binaries consisting of two neutron stars, with masses 
$1.38\,M_\odot$ and $1.42\,M_\odot,$ two black  holes
with masses $4.8\,M_\odot$ and $5.2\,M_\odot,$ two black  
holes with masses $9.5\,M_\odot$ and $10.5\,M_\odot$ 
and, finally, a neutron star and a black hole binary with 
component masses $1.4\,M_\odot$ and $10\,M_\odot.$}
\label{fig:1}
\end{figure*}

We will present the results of our investigation in two complementary
ways. We will first discuss the effectualness of the different PN
families with each other. Such an analysis will help us understand
how well the PN approximation has ``converged'' for the  selection
of detection templates. We then go on to look
at the effectualness of the different approximants with the EOB
signal that contains not only the inspiral but
also the merger and ringdown parts. The goal of the latter analysis is
to identify the region in the parameter space where one can safely 
use any PN approximant template in a search, without worrying about 
the loss in signal-to-noise ratio that might arise due to our lack of knowledge of the
real signal, but without expending undue computational resources.
Outside this region, however, one must use template families that
are calibrated to waveforms obtained from numerical relativity simulations.

\subsection{Mutual effectualness of various PN Inspiral template banks}
\label{sec:convergence}

\begin{figure*}[t]
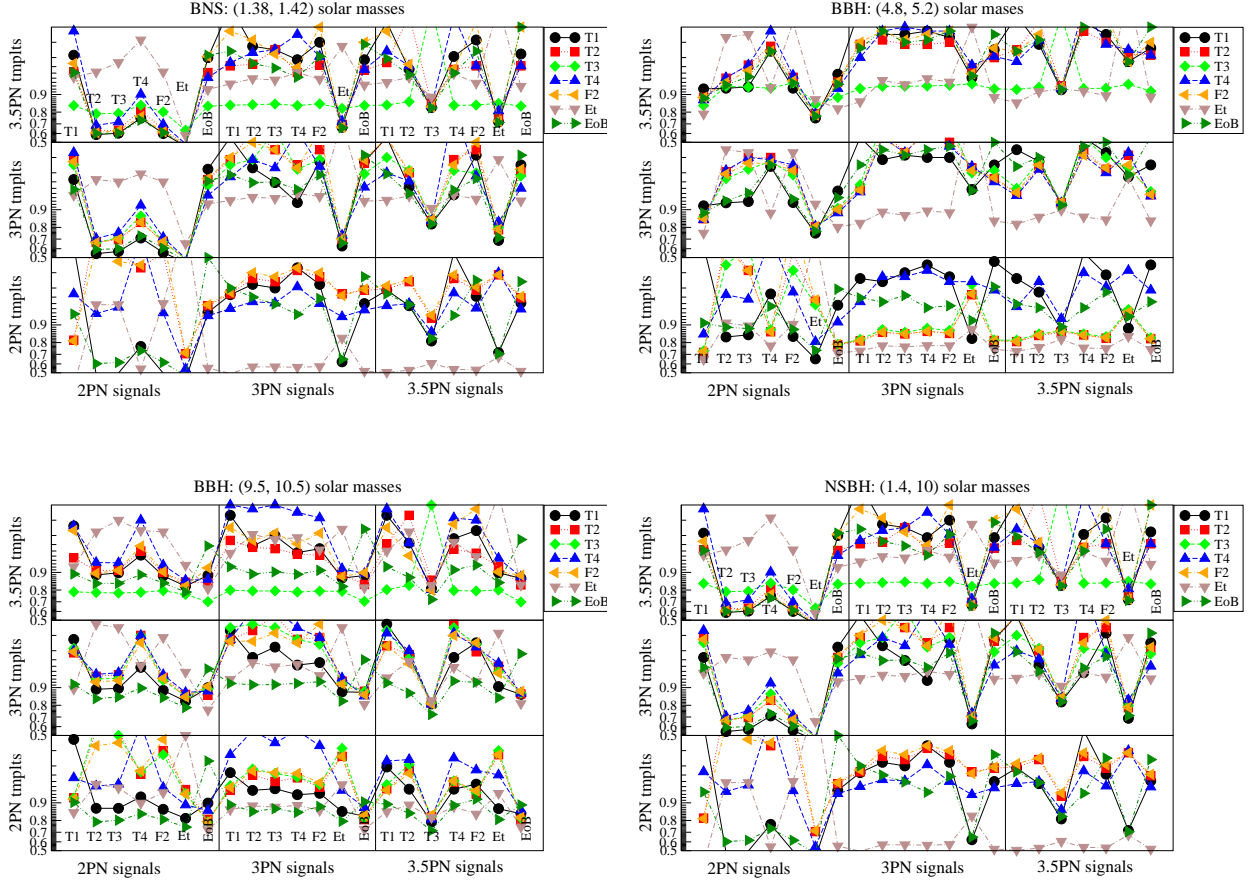

\includegraphics[width=0.45\textwidth]{overlaps/adv-ligo/adv-BNS.eps}
\hskip0.20cm
\includegraphics[width=0.45\textwidth]{overlaps/adv-ligo/adv-BBH2.eps}
\vskip1.0cm
\includegraphics[width=0.45\textwidth]{overlaps/adv-ligo/adv-BBH1.eps}
\hskip0.20cm
\includegraphics[width=0.45\textwidth]{overlaps/adv-ligo/adv-NSBH.eps}
\caption{Same as Fig.\ \ref{fig:1} but for Advanced LIGO.}
\label{fig:2}
\end{figure*}

The effectualness of the different PN families with each other
is shown in Figs.\ \ref{fig:1} (Initial LIGO) and \ref{fig:2} 
(Advanced LIGO) for four different systems with component masses
as indicated at the top of each sub-figure.  In each sub-figure, 
the top panels correspond to the effectualness of different template families at 
3.5PN order, middle panels to 3PN order and bottom panels to 2PN order.
For each template family considered we find their overlap with 
signals from different PN orders (as indicated along the $x$-axis) 
and approximants (as indicated by the text T1, T2, etc.).
Each symbol corresponds to the overlap obtained by a different
template family: (black) circles to TaylorT1, (red) squares to 
TaylorT2, etc., with signals from different PN families. 
Note that we have used the {\it logit scale}\footnote{
Recall ${\rm logit}(p)= \log\left(\frac{p}{1-p}\right).$}
for the vertical axis. This is so that (minor) disagreements 
between the different approximants are made clearly visible.  
Note that since we are considering 
systems with low total mass, say $\leq 20 M_\odot$, 
in this section we use the EOB model terminated at the EOB light 
ring, that is we do not include the merger and ringdown parts.

Conventionally, one says that two approximants $A$ and $B$ are in close agreement with
each other if their mutual effectualness $\epsilon_{AB}$ is 
$0.965$ or greater~\cite {DIS01}. Since in this study we are dealing
with a rather large number of different PN families (21 in all),
we shall relax this condition a bit to $0.95.$ However, we shall
indicate in Sec.\ \ref{sec:region}, the region of the parameter
space where the effectualness is better than $0.965$, but we shall
also quote regions where the effectualness drops to a low value
of $0.9.$ The latter should be helpful for data analysis pipelines 
that  employ a multi-stage hierarchical search, the first stage of
which deploys a coarse grid of templates.  

These figures reveal many different aspects of the (dis)agreements
between the different approximants but we shall only mention
in our discussion the ``diagonal'' behaviour, i.e. overlaps
of each template family with a signal family from the same
PN order.
Focusing first on the Initial LIGO results (Fig.\ \ref{fig:1}), 
we see the evidence for the  clustering of the various approximants 
at 3PN and 3.5PN orders for systems with a smaller total mass. 
In the case of BNS with component masses $(1.38,\, 1.42)\,M_\odot,$ 
2PN ``diagonal'' overlaps are dispersed between
0.74 to 1, 3PN and 3.5PN overlaps are all above 0.95, with TaylorEt 
having the smallest overlaps. 

In the case of BBH with component masses $(4.8,\, 5.2)\,M_\odot$,
2PN overlaps are between 0.8 and 1, 3PN overlaps
are all greater than 0.95 except TaylorEt, 3.5PN overlaps are
greater than 0.95 for all except TaylorEt, TaylorT3 and EOB. 
There are several important points to note: As discussed
in Sec.\ \ref{sec:freq evolution}, TaylorT3 terminates somewhat prematurely
before reaching the last stable orbit. Therefore, one expects 
to have poorer overlaps for all templates if TaylorT3 signal terminates
in band, which will be the case for systems with a total mass greater
than about $10\,M_\odot.$ The asymmetry in the overlaps mentioned
in Sec.\ \ref{sec:maximization} is apparent in the case of TaylorEt: The overlaps of
all templates with TaylorEt signal is greater than the
converse, namely the overlaps of the TaylorEt templates with
other signals. The poorer performance of EOB templates 
(terminated at the light ring) is due to the fact that the 
waveform has power in band beyond the last stable orbit.

In the case of NSBH with component masses $(1.4,\, 10)\,M_\odot,$
2PN ``diagonal'' overlaps are distributed between 0.6 and 1, 3PN and
3.5PN overlaps are consistently above 0.95 except for TaylorEt
signals (both orders) and TaylorT3 (at 3.5PN).

In the case of BBH with component masses $(9.5,\, 10.5)\,M_\odot,$
there is no agreement between approximants irrespective of the PN
order. In this sense, one cannot trust using any particular approximant
as a search template.

Let us now turn to Fig.\ \ref{fig:2} which depicts the results
for Advanced LIGO noise power spectral density.
In the case of BNS with component masses $(1.38,\, 1.42)\,M_\odot,$
the 2PN ``diagonal'' overlaps are between 0.4 and 1 (note that some of
the data points are below the scale of 0.5 that we employ). 
The 3PN (except TaylorEt signal) and 3.5PN (except TaylorT3 template
and TaylorT3 and TaylorEt signals) overlaps are uniformly larger than 0.95.
The effectualness of all templates with TaylorEt signal
is generally  smaller (0.6-0.8) than the effectualness 
with a TaylorEt template.  In the case of BBH with component 
masses $(4.8,\, 5.2)\,M_\odot,$ the 2PN overlaps could be as small 
as 0.65.  At 3PN, all approximants (except TaylorEt templates) 
and 3.5PN (except TaylorEt and TaylorT3 templates) the overlaps are
0.95 or greater.  In the case of NSBH with component masses 
$(1.4,\, 10)\,M_\odot,$ the 2PN overlaps are as low as 0.4.
At 3PN and 3.5PN, the overlaps are larger than 0.95 except in the
case of TaylorEt signals (3PN, 3.5PN) and TaylorT3 templates (3.5PN).
In the case of BBH with component masses $(9.5,\, 10.5)\,M_\odot,$
the 2PN overlap could be as low as 0.7.
The overlaps are larger than 0.95 at 3PN except in the case of EOB 
templates and TaylorEt and EOB signals. Finally, at 3.5PN order
the different approximants are seen not to agree with each other
very well. The cause of these features is the same
as our discussion for Initial LIGO.

\begin{figure*}[t]
\includegraphics[width=0.45\textwidth]{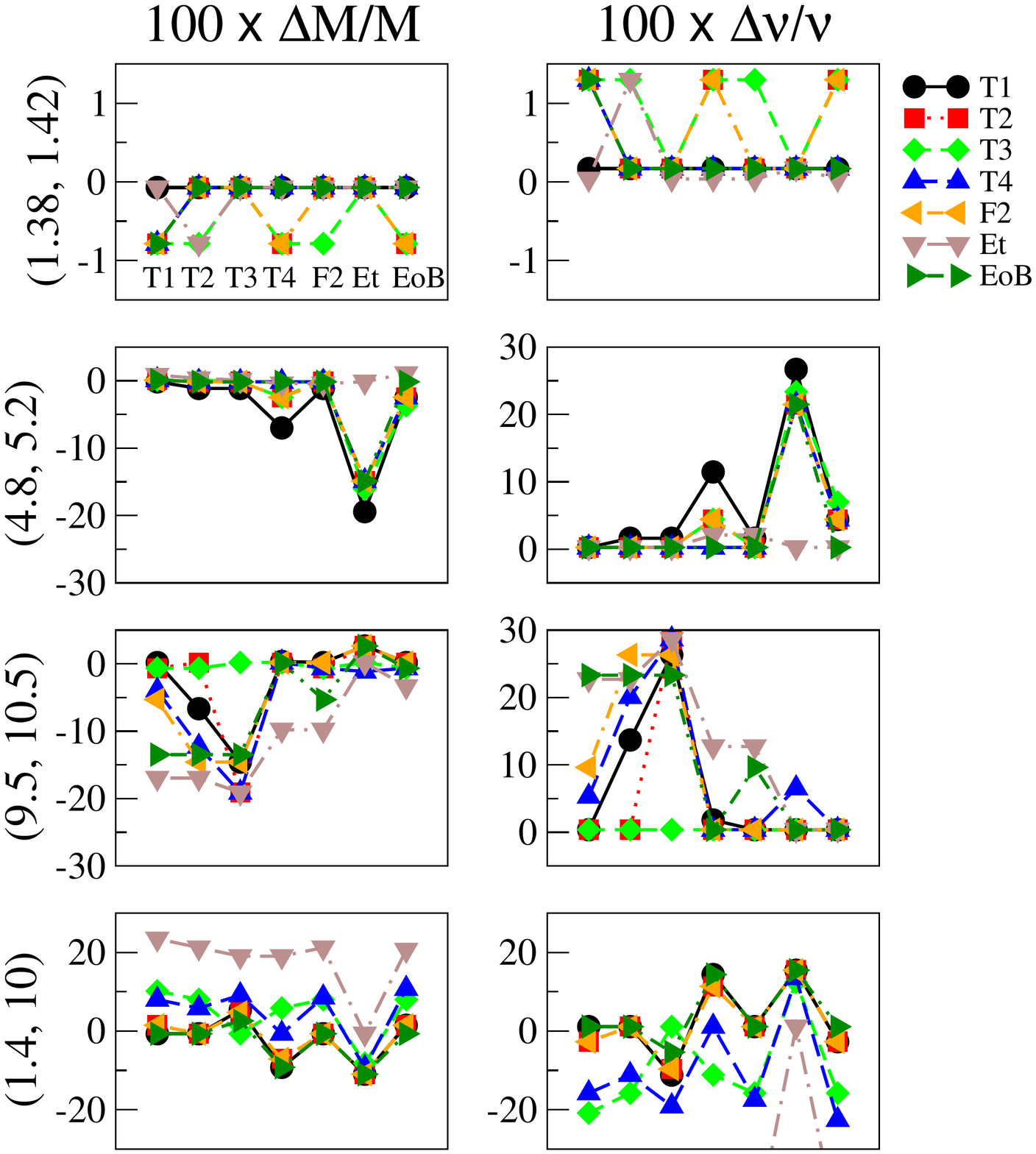}
\includegraphics[width=0.45\textwidth]{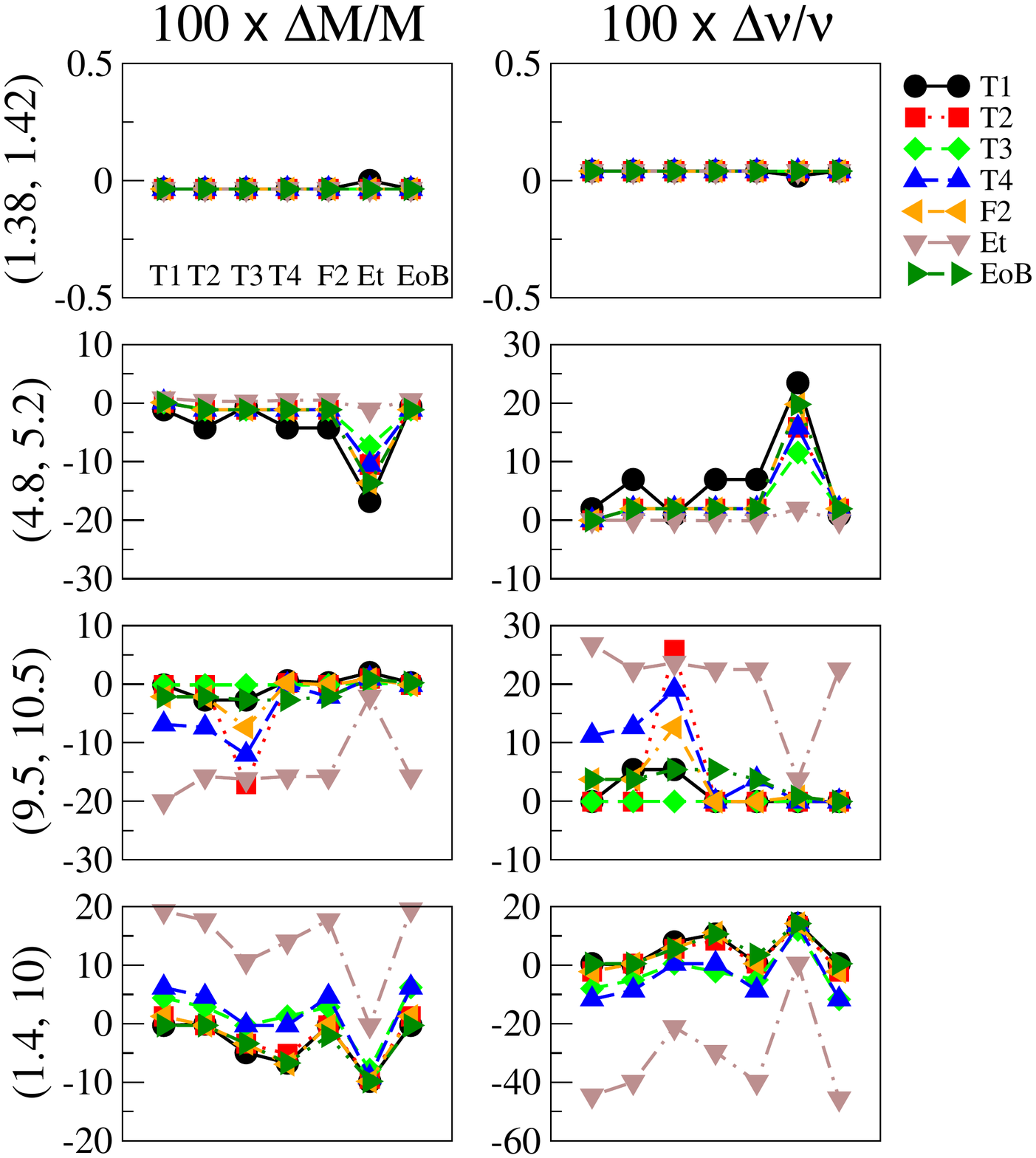}
\caption{Percentage bias in the estimation of the total mass $M$ 
and symmetric mass ratio $\nu$ at 3.5PN order. Left two columns are for Initial LIGO
and the right two for Advanced LIGO.
The bias $\Delta M/M$ is defined as $\Delta M= \left (1-M_{\rm Tmplt}/M_{\rm Sgnl} \right),$
where $M_{\rm Sgnl}$ and $M_{\rm Tmplt}$ denote the total mass corresponding 
to the signal and the template that obtained the maximum effectualness, respectively
(and similarly for $\nu$). What is plotted is percentage bias.
The bias arises because the template family (as indicated in the key) is different from
that of the signal family (as indicated in the top left panel as T1, T2, etc.). 
} 
\label{fig:3}
\end{figure*}

\subsection{Discussion}

In the case of binary neutron stars, the merger 
occurs far outside the sensitive band
of the detector and even the late stages of inspiral is out of
band. Binary neutron stars will very much be in the adiabatic
regime as the signal sweeps through the band and a good test of
the PN approximation is to ask how well the different
waveforms agree with one another in this regime. The finite
bandwidth of the detector essentially probes this regime for
binary neutron stars. Note that the effectualness amongst different PN families at 2PN order
is pretty poor but greater than 0.95 (with the exceptions discussed
earlier) at 3PN and 3.5PN orders.  In the case of Advanced LIGO 
(cf.\ Fig.\ \ref{fig:2}), the lower frequency cutoff used in computing
the overlap integrals is 20 Hz and a binary neutron star spends
more than 750 cycles in band. Effectualness of 0.95 or greater means 
that the waveforms remain in phase over the entire duration
of the signal. Of course, in reality the parameters of the
signal and the template are not the same, but even so this is
a remarkable success of the PN scheme.

\begin{figure*}[t]
\includegraphics[angle=-90,width=0.40\textwidth]{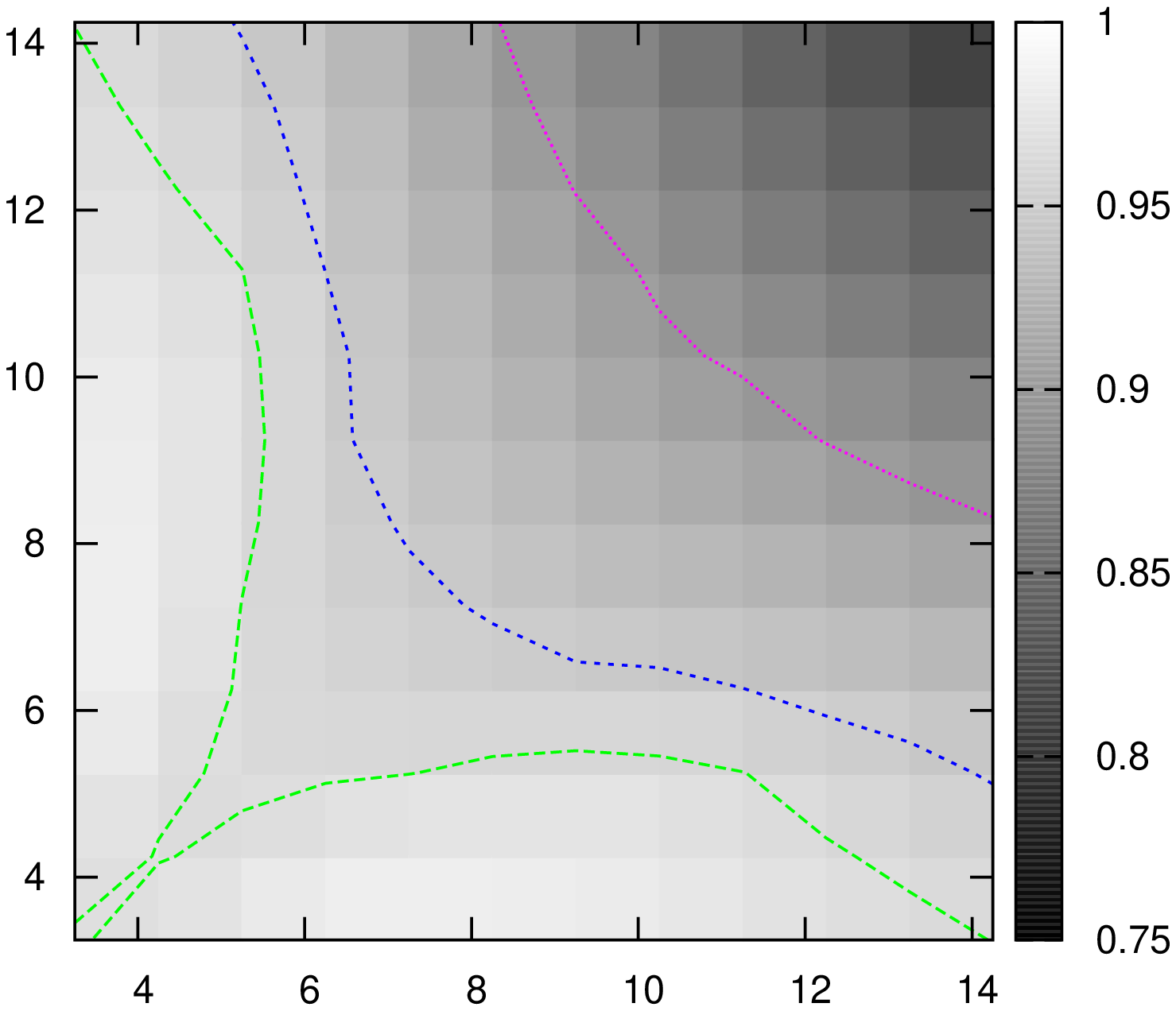}
\hskip -2.6 cm
\includegraphics[angle=-90,width=0.40\textwidth]{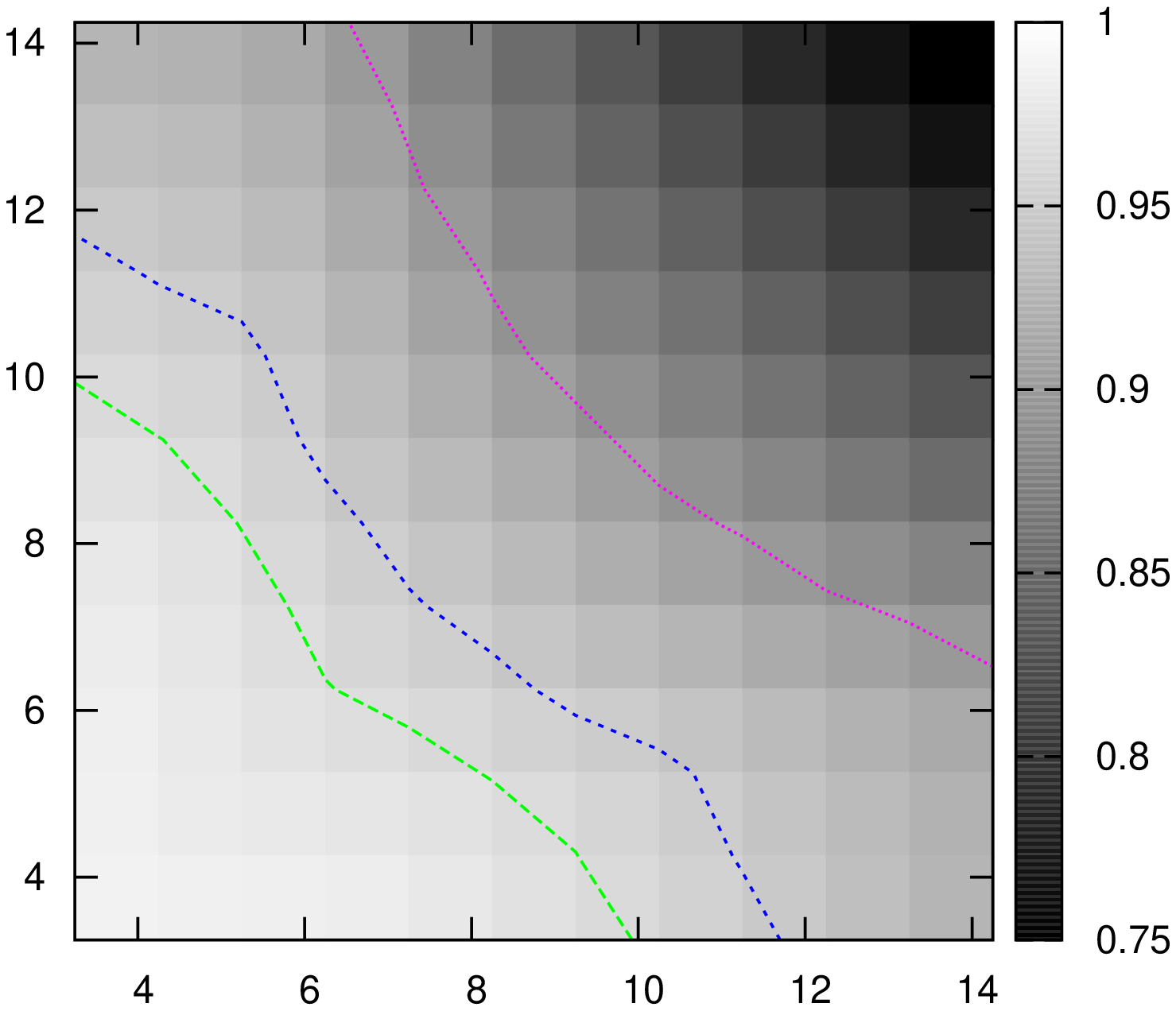}
\hskip -2.6 cm
\includegraphics[angle=-90,width=0.40\textwidth]{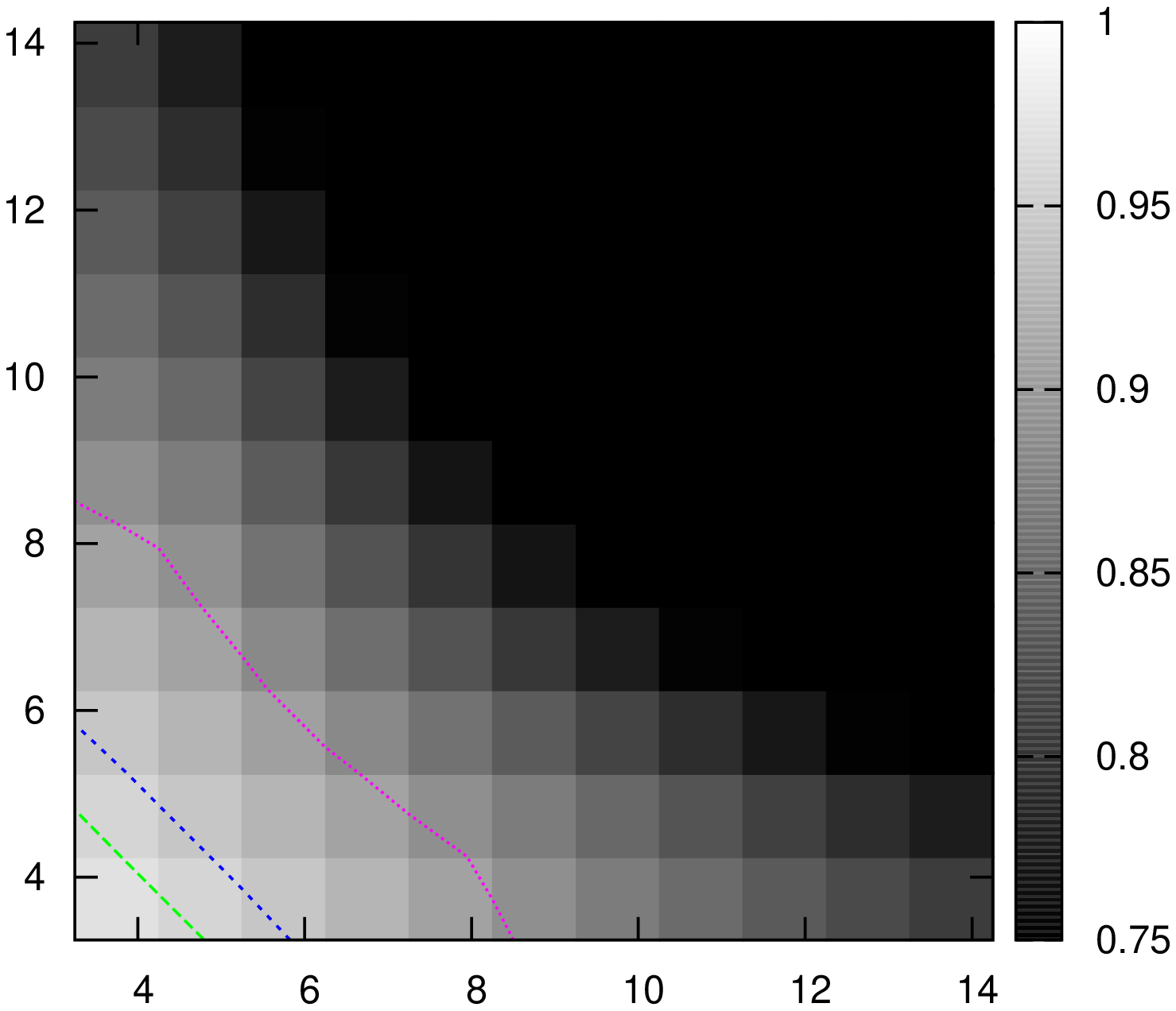}
\vskip-1.75 cm
\includegraphics[angle=-90,width=0.40\textwidth]{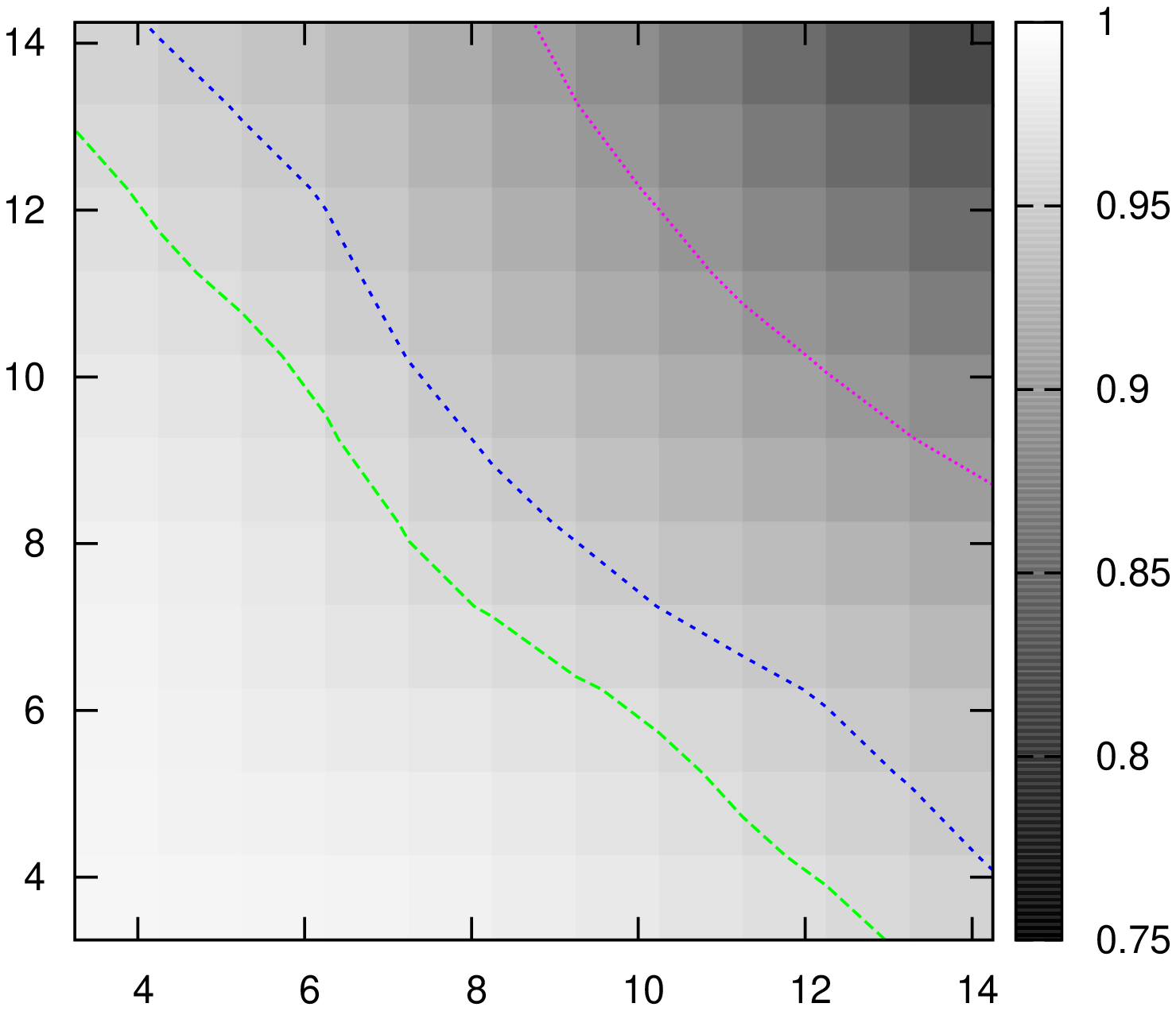}
\hskip -2.6 cm 
\includegraphics[angle=-90,width=0.40\textwidth]{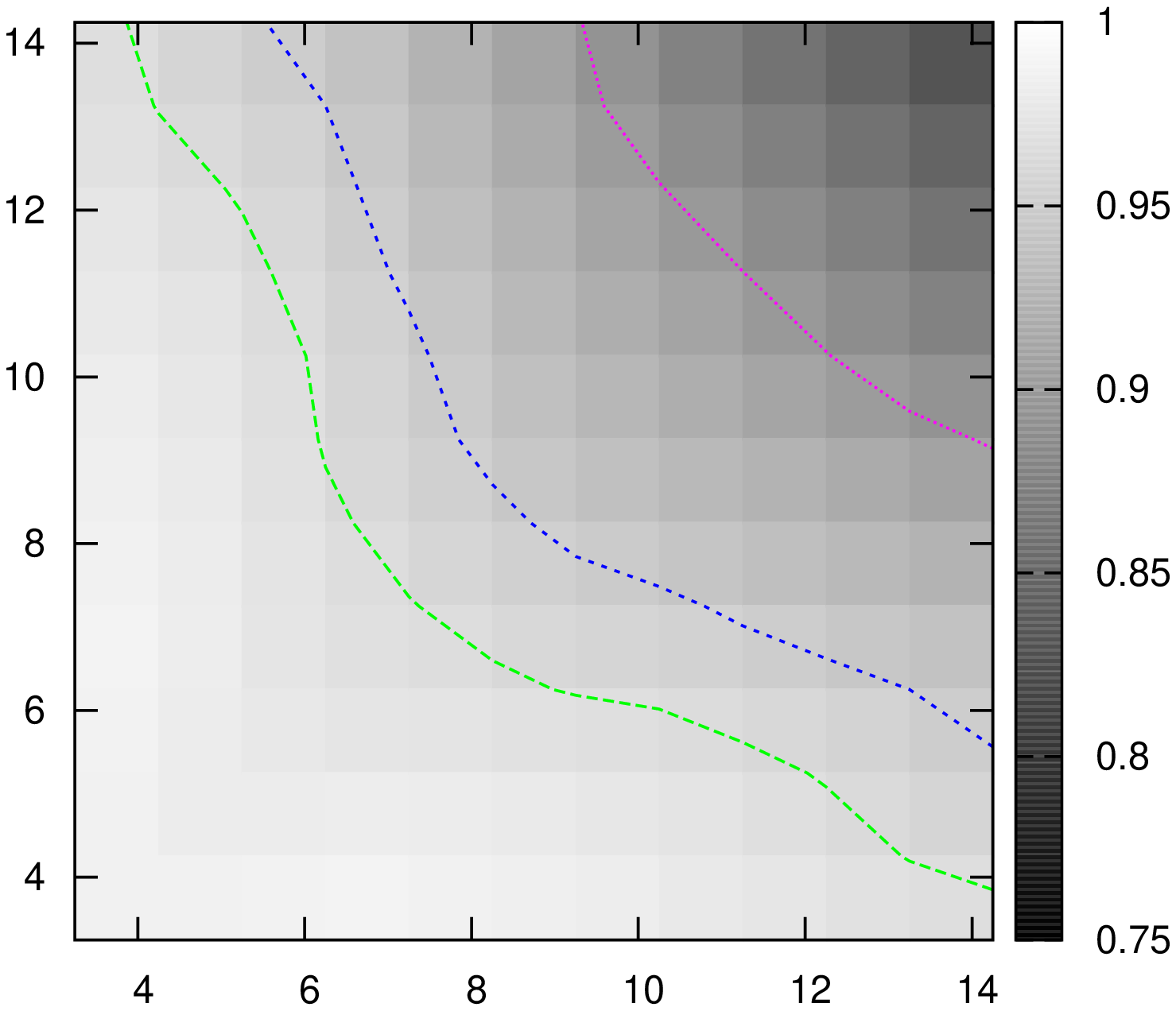}
\hskip -2.6 cm 
\includegraphics[angle=-90,width=0.40\textwidth]{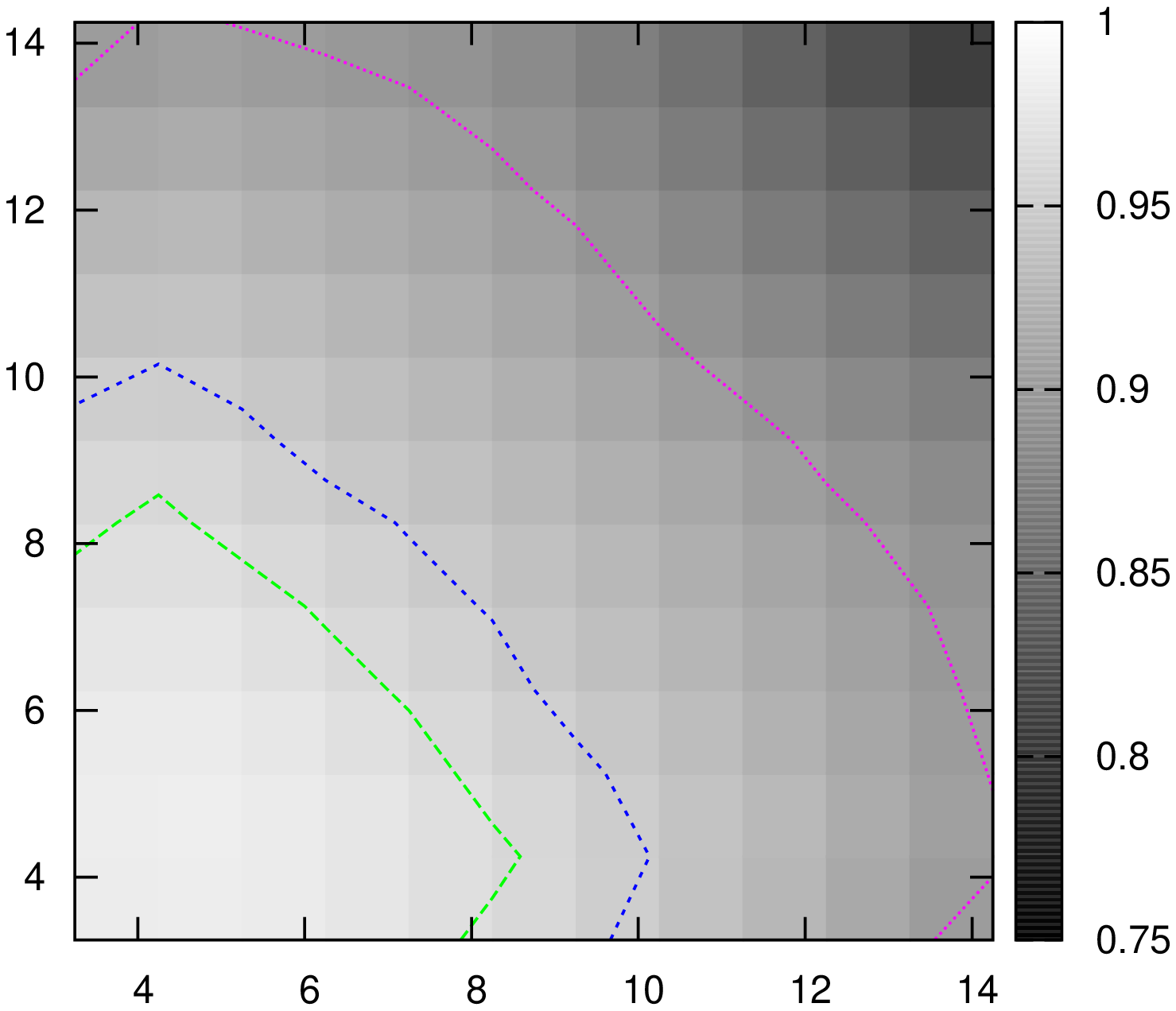}
\caption{Overlaps of different 3.5PN approximants   with the EOB 
inspiral-merger-ringdown signal in Initial LIGO in the $(m_1-m_2)M_\odot$
plane. The approximants considered 
from left-to-right are TaylorT1, TaylorT2, TaylorT3 (top panels), and
TaylorT4, TaylorF2, TaylorEt (bottom panels).
The contours correspond to overlaps of 0.9, 0.95 and 0.965.}
\label{fig:overlaps ini-ligo}
\end{figure*}

For a BBH system with masses $(4.8,\, 5.2)\, M_\odot,$ 
we see that 2PN and 3PN order templates are qualitatively
similar to the binary neutron star case. However, we can see
a marked deterioration of the effectualness at 3.5PN order. For 
a system of total mass of $10\,M_\odot,$ the Schwarzschild 
LSO occurs at $\sim 440$ Hz and the detector is sensitive to 
the late stages of the inspiral phase. It is not entirely 
surprising, therefore, that different PN orders do not agree with each
other to the same extent as in the binary neutron star case.
However, note that, with the exception of TaylorT3, which
terminates at a frequency somewhat lower than others, and
TaylorEt, all other templates have effectualness of 0.95 
or better with each other. Among approximants that agree
with each other, EOB has the smallest effectualness. This is
because the latter model contains the plunge phase of the 
coalescence with ending frequencies far higher than the LSO
while other approximants do not have the plunge phase.

The LSO of a BBH with component masses $(9.5,\, 10.5)\, M_\odot,$ 
is at $\sim 220$ Hz and the plunge phase spans $220$ Hz to
about $600$ Hz. Therefore, the detector is pretty sensitive
to the late phases of the coalescence. We see 
deterioration of the effectualness, both at 3PN and 3.5PN 
orders.  Apart from TaylorT3, whose poor overlaps at 3.5PN are 
explained by the early termination of the signal, the EOB 
stands out by achieving overlaps as low as 0.92 with other
families. 

As a final example, the effectualness of templates for a signal from 
a neutron star-black hole binary of masses $(1.4,\, 10)\, M_\odot,$
we see that the different PN families, including the EOB, 
are in good agreement with each other, with the sole exception 
of TaylorEt.  In fact, the convergence amongst different families 
seems to be somewhat better than the BBH system of component
masses $(9.5,\,10.5)\,M_\odot.$

At this juncture, it is worth pointing out that our numerical results
for effectualness in the subset of cases where TaylorEt is chosen as
 the signal model, are consistent with those in Ref.~\cite{Bose:2008ix}, 
which investigated the {\it fitting factors} to ascertain if 
3.5PN TaylorEt {\it signals} could be effectually and faithfully 
searched by TaylorT1, TaylorT4 and TaylorF2 templates. 
There is agreement too on the general features of our results
with regard to systematic biases, the dependence on the total mass 
and  qualitative factors underlying them.  However, this agreement 
of numerical results for faithfulness and effectualness
in no way  extends to the general motivation and claims regarding
the TaylorEt approximants
~\cite{Gopakumar:2007jz,Bose:2008ix,Tessmer:2008tq}
and, hence, are worth clarifying.

Indeed, there is no basis to refer to the $x$-based orbital phasing equation
Eq.~(\ref{eq:phasing formula1a}) as Newtonian~\cite{Tessmer:2008tq}, since 
the $\omega$ here is $n$PN  accurate (depending on the PN-generation order 
one is working at) and {\it implicitly} incorporates conservative contributions 
to gravitational-wave phase evolution at various PN orders.  
It is incorrect to claim~\cite{Bose:2008ix} that
conservative contributions to the gravitational-wave phase evolution do not
appear in the standard approximants, or that the TaylorEt-based scheme 
treats conservative and radiation-reaction contributions more equitably 
than the standard $x$-based approximants.
It is misleading~\cite{Bose:2008ix} to refer  to {\it only} 
TaylorEt-based approximants
as ``fully gauge invariant in contrast to EOB'' (especially
in the circular orbit case).
All $x$-based schemes are also fully gauge invariant.
Finally, one may work in specific convenient coordinate systems as
do EOB and numerical relativity  simulations, as long as
one deals  with and compares gauge invariant quantities
at the end.

In our view,
 the very different behaviour of the TaylorEt approximant relative to the
standard $x$-based approximants may be traced to the manner in which
the orbital phasing is ``packaged'' in the two schemes. In the $x$-based schemes
the orbital phasing is implicitly in a resummed form, since the phasing
is written in an appropriate PN-accurate angular velocity $\omega_{n\rm PN}$
($n=2,3$ for 2PN, 3PN templates). On the other hand, the representation
in terms of $\zeta,$ relative to the $x$ schemes, is a re-expanded form.
And indeed, based on the comparison between analytical schemes
and numerical relativity simulations, the $\zeta$ schemes do relatively
worse. The feature related to the monotonic-convergence of the TaylorEt scheme
is of secondary importance in comparison to the main requirement
of high  phasing accuracy of an analytical model with numerical relativity
simulations over all mass-ratios.

A few general comments are in order before we conclude this Section. 
We do not believe that at present there are convincing theoretical reasons to consider any one
particular PN family of inspiral models to be a privileged signal model.
Consequently, the best that one can do is to examine the mutual closeness
of these various inspiral models, as we have done, and work at the PN
order where these various template families display the greatest agreement.
It is precisely in this regard that the viewpoint we present here
differs  from those in~\cite{Gopakumar:2007jz,Bose:2008ix,Tessmer:2008tq}
which assumes primacy for one specific approximant, namely the TaylorEt
approximant, based on theoretical motivations that at present do not 
appear to be fundamental or compelling.
Consequently, though there is no difference in the numerical results
in the subset of cases that are common in our investigations, 
there is a big difference
in the conclusions that we believe can be inferred.
For instance, before one can legitimately decide on the inability of standard
template banks in the gravitational data pipeline to detect signals 
from binaries with eccentricity~\cite{Tessmer:2008tq}, 
it is necessary to first fold in the differences in the simpler 
quasi-circular case arising on account of different parametrisations. 
Similar considerations should be borne in mind when dealing with
analogous problems in the  spinning case.
 
Based on the analysis presented heretofore, we conclude 
that the PN approximation has pretty much converged at
3PN and 3.5PN orders\footnote{Though qualitatively  we may expect 
similar results for Virgo, quantification requires an 
analysis using the Virgo noise curves. Needless to add,
that the situation for a space detector like LISA can be 
expected to be even more different and interesting
to study}, as long as the total mass is less than about $12\, M_\odot$
(with the exceptions discussed in the previous Section).

For heavier binaries, the approximants begin
to differ considerably, and this is almost entirely 
because the adiabatic approximation begins to breakdown 
and the plunge and the merger phases become more and more 
important.  Hence, in the next Section we will supplement  
the present analysis by looking more precisely into the overlaps of 
the different PN templates  with a prototype of the more complete
signal model, namely the EOB model, including the merger and ringdown parts.

\begin{figure*}[t]
\includegraphics[angle=-90,width=0.40\textwidth]{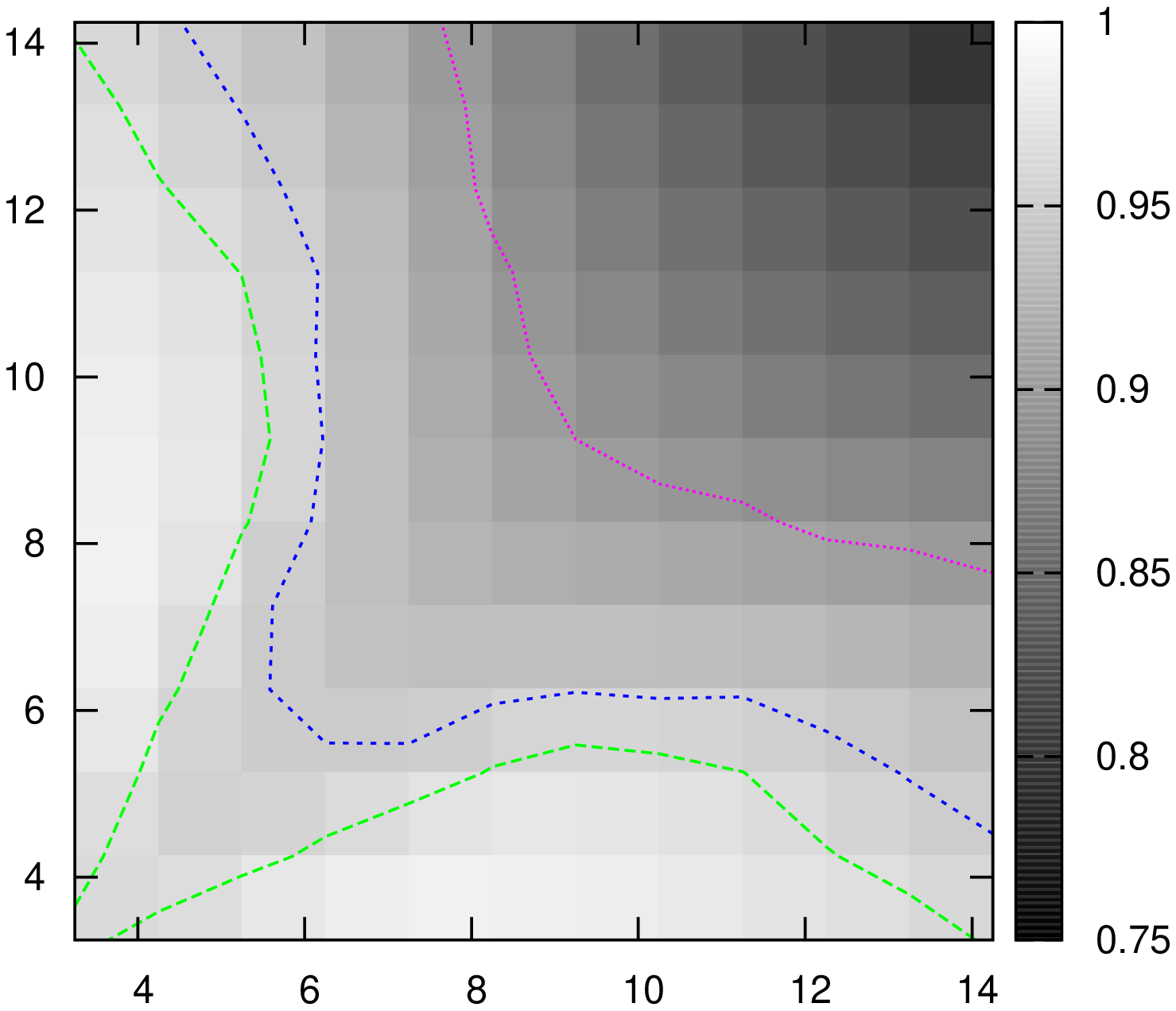}
\hskip -2.6 cm 
\includegraphics[angle=-90,width=0.40\textwidth]{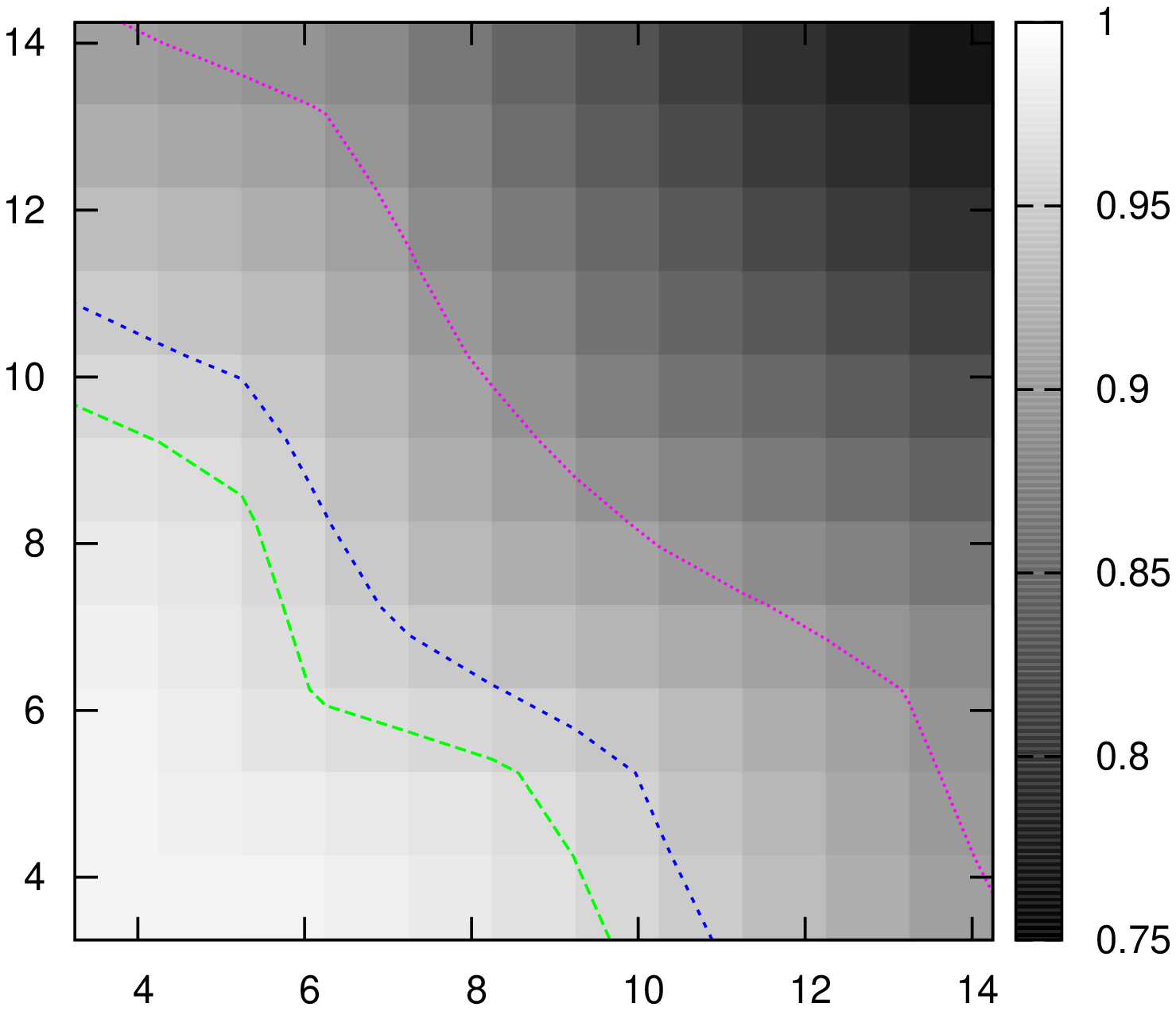}
\hskip -2.6 cm 
\includegraphics[angle=-90,width=0.40\textwidth]{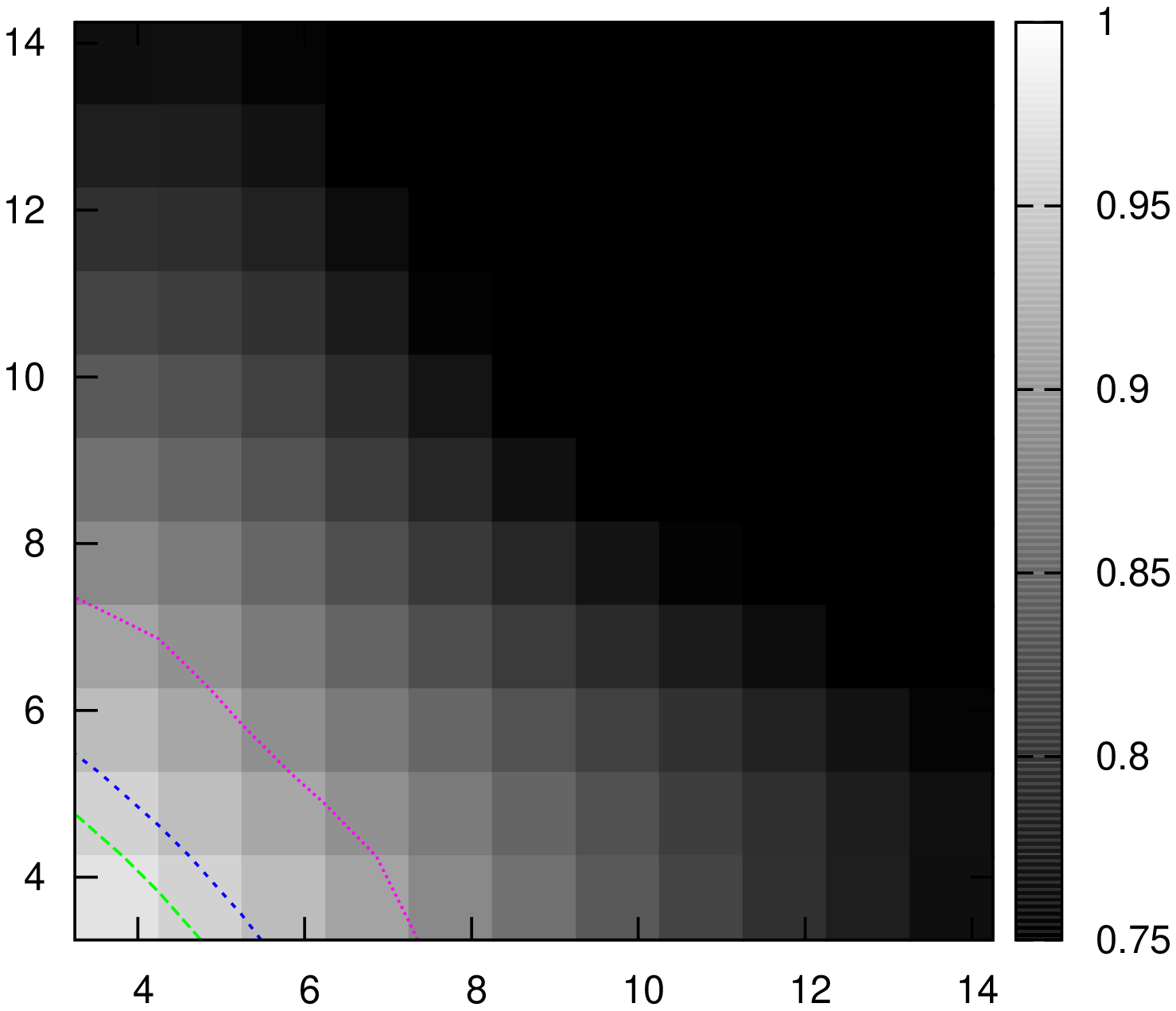}
\vskip-1.75 cm
\includegraphics[angle=-90,width=0.40\textwidth]{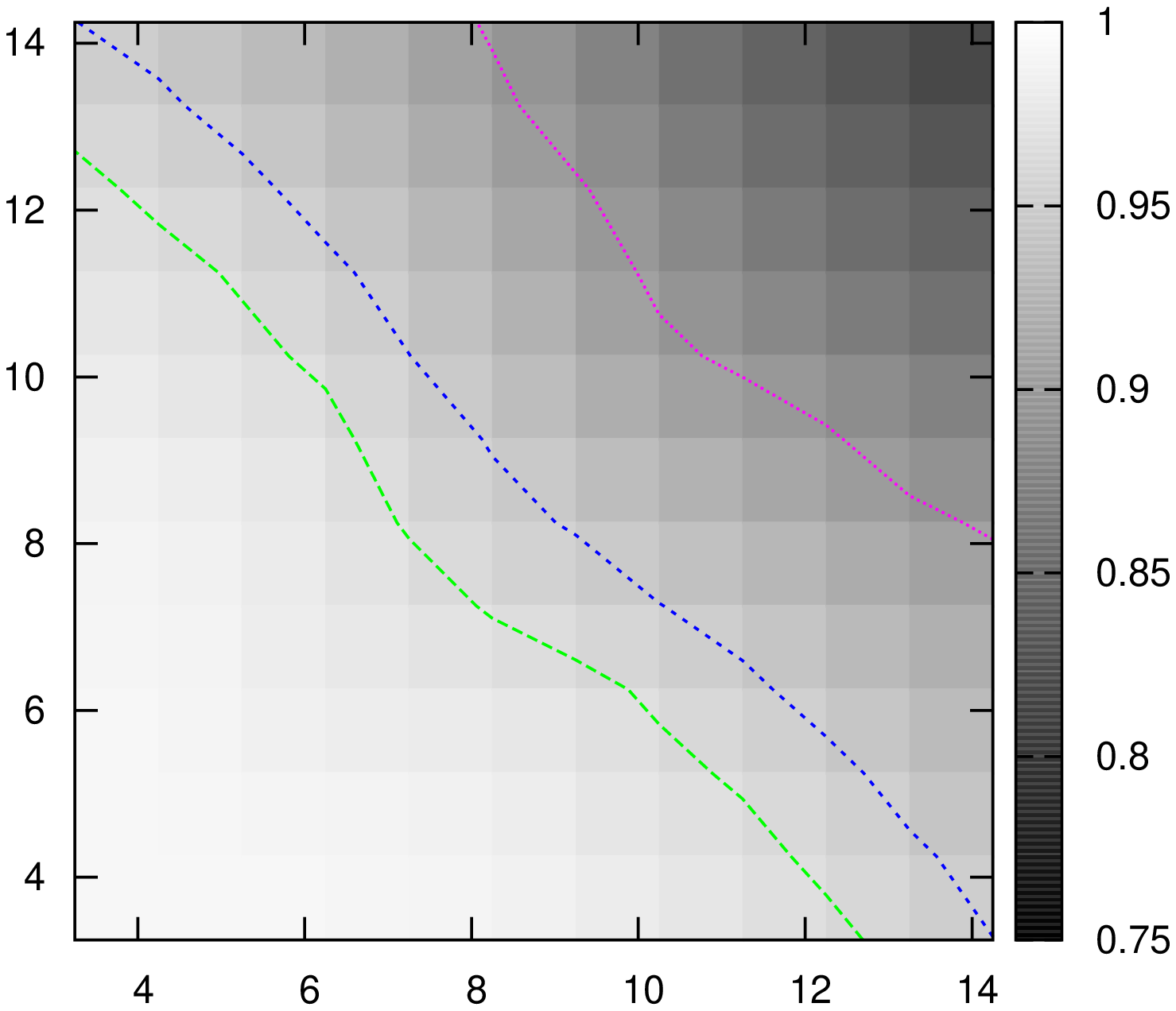}
\hskip -2.6 cm 
\includegraphics[angle=-90,width=0.40\textwidth]{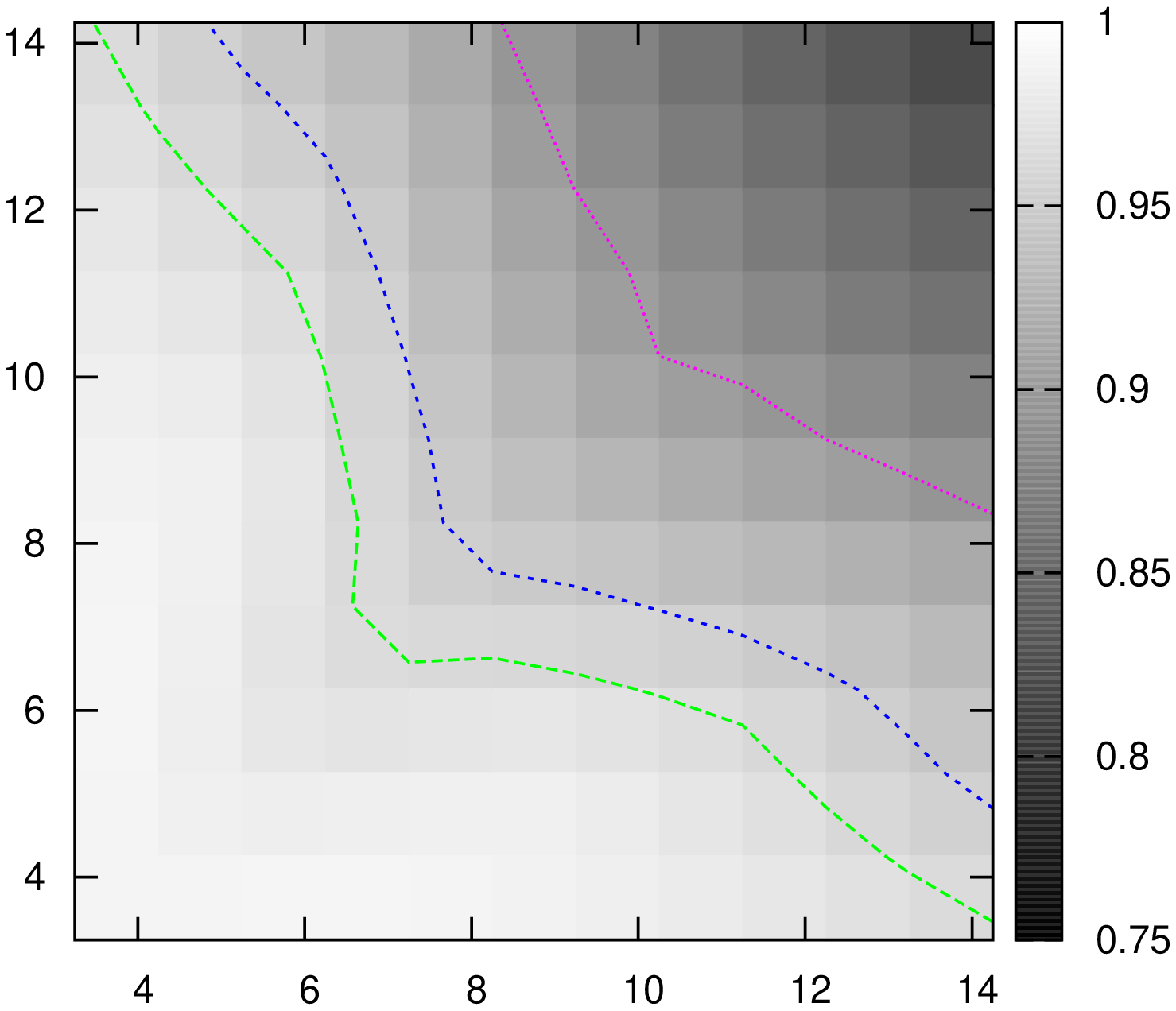}
\hskip -2.6 cm 
\includegraphics[angle=-90,width=0.40\textwidth]{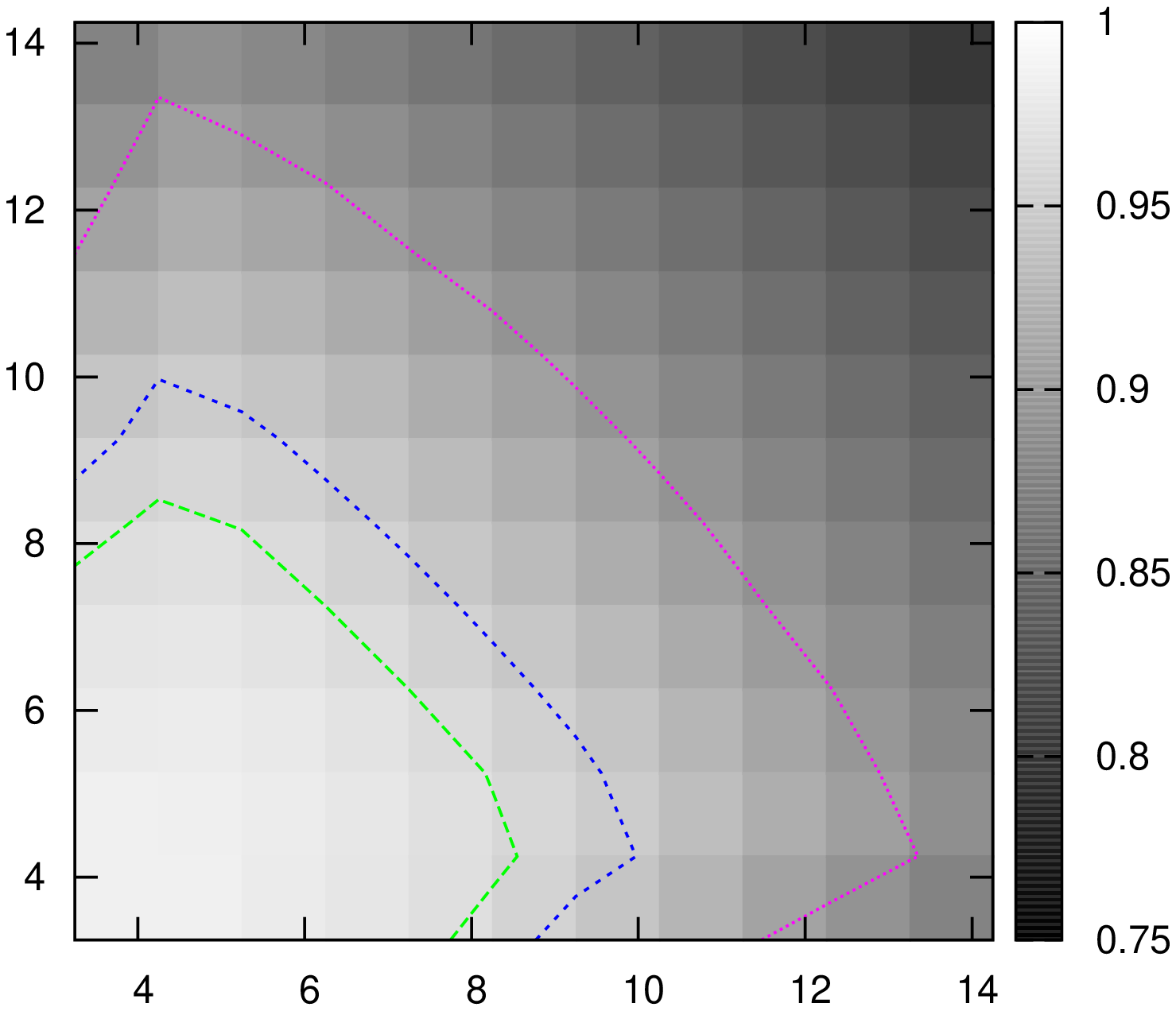}
\caption{Same as Fig.\ \ref{fig:overlaps ini-ligo} except that the noise
spectral density is that of Advanced LIGO. The contours correspond to overlaps 
of 0.9, 0.95 and 0.965.}
\label{fig:overlaps adv-ligo}
\end{figure*}

\subsection{Biases in the estimation of parameters}

Recall that, in the computation of the effectualness one 
maximizes the scalar product of a (normalized) signal 
with a template over the parameters of the template
keeping those of the signal fixed. Therefore, one can
get an idea of how dissimilar the parameters of an
approximant need to be in order to match a given signal.
This is a systematic effect that leads to a bias in the
estimation of parameters if the template approximant is not
the same as the signal approximant. Let the total mass
of the signal and template waveforms be, respectively, 
$M_{\rm Sgnl}$ and $M_{\rm Tmplt},$ when the scalar
product is maximized. The percentage bias $\Delta M$ in
the total mass is defined as $\Delta M= 100( 1 - M_{\rm Tmplt}
/M_{\rm Sgnl}),$ and similarly for the symmetric mass ratio
$\nu.$

For a given binary, the biases are qualitatively similar 
for Initial and Advanced LIGO noise power spectral densities. 
In general, the biases are appreciably smaller at 3PN and 3.5PN 
orders than at 2PN order and progressively increase with the total mass,
although they are far larger than the statistical errors computed
using the Fisher information matrix~\cite{Arun:2004hn}.
Figs.\ \ref{fig:3} plots the percentage biases
in the total mass $M$ and symmetric mass ratio $\nu$ at 3.5PN order.
The left two (right two) columns use the Intial LIGO (Advanced LIGO) noise 
spectral density.
For the four systems considered, namely $(1.38\,, 1.42)\, M_\odot,$
$(4.5\,, 5.2)\, M_\odot,$ $(1.4\,, 10)\, M_\odot,$ and
$(9.5\,, 10.5)\, M_\odot$ binaries, the largest bias in 
the total mass $M$ is 1\%, 20\%, 20\% and 20\%, respectively, and
the symmetric mass ratio $\nu$ is 1\%, 25\%, 70\% and
25\%, respectively.

\section{Results of the effectualness of PN templates with the full waveform}
\label{sec:region}

Having established the convergence of PN approximations at 3PN and
3.5PN orders (for determining effectual templates for detection)
in the regime where the approximation is expected to
be valid, let us now examine the region in the parameter space where
PN families can be used as search templates. To achieve this
goal we will use the EOB model calibrated to numerical relativity
simulations~\cite{Buonanno:2007pf}. For brevity, we have omitted 
plots of the effectualness of the 3PN approximants with this EOB model; 
they are quite similar to the 3.5PN plots.

Although Ref.\ \cite{Buonanno:2007pf} explored the agreement
between the EOB model and numerical simulations for several 
modes, in this study we will work with only the dominant harmonic (i.e., the 
$h_{22}$ mode) at leading PN order. Higher-order 
amplitude corrections are known to be important for parameter
estimation~\cite{VanDenBroeck:2006ar,VanDenBroeck:2006qu} and a 
future study must repeat this investigation with the full waveforms.

Fig.\ \ref{fig:overlaps ini-ligo} shows the effectualness of the six
PN families TaylorT1, TaylorT2, TaylorT3 (top panels, respectively
from left to right), TaylorT4, TaylorF2, and TaylorEt (bottom panels, 
respectively from left to right) for Initial LIGO noise power 
spectral density.  Fig.\ \ref{fig:overlaps adv-ligo} 
shows the same but for Advanced LIGO noise power spectral density.
The effectualness was computed using a hexagonal template bank 
~\cite{Cokelaer:2007kx} and is shown as a gray-scale map in the space of 
the component masses that are taken to vary from $3\, M_\odot$ to 
$14.5\, M_\odot.$ For all the maps we have chosen the gray-scale 
to vary from $0.76$ to $1.$ The dotted contours show effectualness 
at three values: $0.965,$ $0.95$ and $0.90.$ 

The trends of the overlaps is rather similar irrespective of which
noise power spectral density we use, although the actual overlaps
are systematically smaller in the case of Advanced LIGO as compared
to Initial LIGO. This is due to the broader frequency sensitivity
of the former in relation to the latter. The following discussion
is, therefore, applicable in both cases.

Let us first note some peculiarities. TaylorT3 at 3.5PN 
leads to particularly ineffectual templates. As mentioned before,
TaylorT3 at 3.5PN terminates rather prematurely. The LSO defined by
the Schwarzschild potential is at $f_{\rm LSO} \sim (440/10\,M_\odot)\,
\rm Hz,$ 
but TaylorT3 at 3.5PN approximants terminate at 
$\sim (220/10\, M_\odot)\,\rm  Hz.$
This discrepancy is so large that even with 
the biases in the component masses allowed in the computation 
of the effectualness (recall that we maximize the overlap over 
template masses), which, in principle, makes it possible for a 
template of a lower mass to match a signal of a higher mass, TaylorT3 
is unable to achieve good overlaps. This is because a mismatch in
the component masses can make a template more, or less, asymmetric
than the signal, which has the effect of increasing, or decreasing,
the duration of the template relative to the signal. While small
differences in the ending frequencies can be achieved by a mismatch
in the total mass without affecting the signal duration too greatly,
large differences cannot be compensated by such a mismatch in the
parameters.

At 3PN and 3.5PN the effectualness of TaylorEt with a EOB signal
for a binary of component masses $(3,\,10)\,M_\odot$ 
[respectively, $(10,\,10)\,M_\odot$] is 0.83 and 0.90 
[respectively, 0.87 and 0.89]. This is because amongst all PN 
approximants TaylorEt seems to converge far slower than any other. 
Further, an examination of the coefficients in the PN 
terms of the phasing formulas in Eqs.\ (\ref{eq:Et}) indicates 
that higher order PN terms have increasingly greater coefficients. 
In general,  it has been observed that 
the appearance  of such larger coefficients in higher
order terms of an approximant scheme
inevitably worsens its convergence and the present instance may be no
exception to this case\footnote{While comparing the coefficients 
it may be useful to note that $v\simeq 1/\sqrt{6}$ corresponds to 
$\zeta$ in the range of $0.136$-$0.138$ depending on the symmetric 
mass ratio $\nu$ and the PN order.}.

With the exception of the peculiarities noted above, we see 
that all approximants do progressively better at higher PN orders.
Conclusions drawn in the previous Section with regard to the convergence
of the PN approximations are further corroborated here where
we have measured the overlaps with a signal that is matched to 
numerical relativity simulation, which can, therefore, be taken to 
be close to what a real signal might be. 

Computationally, TaylorF2, with its phasing formulas given explicitly 
in the Fourier domain, is the least expensive. This is because 
matched filtering is most easily  carried out in the Fourier domain, 
which means that a time-domain approximant must be Fourier transformed 
before computing the cross correlation.
By employing TaylorF2 models one can avoid one forward Fourier transform.
Moreover, TaylorF2 offers the flexibility in the choice of the ending
frequency. Unlike the time-domain models, which have either a natural 
ending frequency defined by the extremum 
of the binding energy or the frequency evolution stops before reaching
LSO, TaylorF2 has no such restriction. In fact, as obtained in Refs.~\cite{Pan:2007nw,Boyle:2009dg}, 
by extending the upper cutoff beyond the usual upper cutoff (i.e., the Schwarzschild LSO), 
the TaylorF2 model matches remarkably well with numerical relativity 
waveforms for a far greater range of masses. However, as noted in Ref.~\ \cite{Boyle:2009dg} the ending frequency that
must be employed in order to achieve the best match with numerical-relativity
 waveforms depends on the noise power spectral density.
 This could turn out to be 
an unnecessary computational burden
in a data analysis pipeline. The alternative is to choose the upper 
frequency cutoff as an additional search parameter or allow unphysical 
values of $\nu > 0.25$~\cite{Pan:2007nw,Bose:2008ix,Boyle:2009dg} or 
to include a p4PN term in the template phase and calibrate it to 
numerical simulations~\cite{Pan:2007nw}. The first two choices would 
result in an unwarranted increase in the computational cost of 
a search as also in the false alarm rate, and we advice against it. The 
third choice could be pursued, but it should be augmented by a more complete 
description of the merger/ringdown signal --- for example by introducing 
a slope break in the waveform amplitude and a superposition of 
Lorentzians~\cite{Pan:2007nw,Ajith:2007kx} 

If a search requires the minimal match to be much smaller than $0.95$
(as, for example, in a hierarchical search) one can extend a search
with TaylorF2 to a total mass of $20\, M_\odot$ with effectualness of 
$0.90.$ 

Before advanced detectors begin to operate, there will be a period
when LIGO and Virgo will operate with sensitivities slightly larger
than, but bandwidths similar to initial detectors -- the so-called
Enhanced LIGO and Virgo+. Since Virgo and Virgo+ are expected to have 
a sensitivity bandwidth similar to Advanced LIGO the results presented 
in this paper are qualitatively similar to in those cases too. Moreover, 
as our results are only sensitive to the bandwidth, conclusions drawn 
by using the noise spectral density of Initial LIGO will also be
valid for Enhanced LIGO. 

All approximants (no exceptions)
achieve an effectualness of $0.95$ or better at 3PN and 3.5PN orders, 
for binaries whose total mass is less than about $\sim 12\, M_\odot.$ 
From the view point of effectualness alone, we conclude that searches for 
binary black holes, in Initial, Enhanced and Advanced LIGO, could 
employ any of the 3PN or 3.5PN families as long as the total 
mass is smaller than about $\sim 12\, M_\odot.$ The final choice of 
the PN family should be based on other criteria. If it is desired that
the minimal match of a template bank is 0.965 or greater, then the
best strategy would be to use the full EOB waveform calibrated 
to numerical relativity.

Another criteria to be considered is the computational cost. A 
typical matched filter search in LIGO data must compute thousands of 
template signals for every 2048 second data segment. 
This can be a heavy burden if 
it takes a significant amount of time to compute each template. 
The EOB templates are computed in the time domain by solving a set of 
differential equations, and the frequency domain signal is then computed via 
Fourier transform. For low-mass systems this cost can become significant 
and will of course vary depending on the implementation and hardware used. 

We have estimated the cost to compute TaylorF2 and EOB templates 
using their implementation in the LIGO Algorithm Libraries (LAL) 
code used for matched filtering searches in LIGO data. We find that for a 
total mass $\geq 40\,M_\odot$, the EOB templates take a factor of 2 longer to 
generate than the same TaylorF2 signals. For a $(10,\,10)\,M_\odot$, $(5,\,5)\,M_\odot$ 
and $(1.4,\,1.4)\,M_\odot$ binary, the EOB templates take 
about a factor of 3, 7 and 20, longer to generate, respectively.  
We tested the waveform generation on a high performance computer with 
32 2.7 GHz CPUs and 132 GB of RAM. On this system, 
EOB templates with a total mass $\geq 40\,M_\odot$ can be generated 
in about $0.1$s, while the $(10,\,10)\,M_\odot$ EOB template 
could be generated in about $0.5$s. 
Since LIGO searches employ thousands of 
CPUs, this is feasible. However, for lower mass signals, the time needed grows 
rather quickly and about $4$s are needed to compute the $(1.4,\,1.4)\,M_\odot$ 
EOB template. 
It may be possible to reduce the computational cost somewhat 
by optimizing the EOB waveform generation code, but the lowest mass templates 
would almost certainly still have a significant computational cost. 
Thus, the increased computational cost must be weighed 
against the benefit of increased effectualness for lower mass signals.

\section{Faithfulness}
\label{sec:faithfulness}

\begin{table*}[t]
\caption{Faithfulness of different approximants for $(1.42,1.38)\,M_\odot$
(left panel) and $(5.2,4.8)\,M_\odot$ (right panel) binaries. 
The rows label template approximant, while the columns label 
signal approximant. For each pair, the top number is Initial LIGO while 
the bottom number is Advanced LIGO. All approximants are at 3.5PN order, 
except our EOB model which has a p4PN coefficient.}
\label{tab:faithful_I}
\vspace{0.5cm}
\begin{minipage}[b]{0.45\linewidth}
\centering
\begin{tabular}{cccccccc}
\hline
\hline
  & EOB & T1   & T2   & T3   & T4   & Et   & F2   \\
\hline
EOB & 1 & .969 & .994 & .997 & .990 & .970 & .994 \\ 
    & 1 & .971 & .996 & .998 & .991 & .974 & .996 \\ \hline
T1  & .969 & 1 & .982 & .981 & .987 & .928 & .982 \\
    & .971 & 1 & .984 & .983 & .990 & .920 & .984 \\ \hline
T2  & .994 & .982 & 1 & .998 & .999 & .958 & 1.000 \\
    & .996 & .984 & 1 & .999 & .999 & .961 & 1.000 \\ \hline
T3  & .997 & .981 & .998 & 1 & .997 & .959 & .998 \\
    & .998 & .983 & .999 & 1 & .998 & .961 & .999 \\ \hline
T4  & .990 & .987 & .999 & .997 & 1 & .950 & .999 \\
    & .991 & .990 & .999 & .998 & 1 & .949 & .999 \\ \hline
Et  & .970 & .928 & .958 & .959 & .950 & 1 & .958 \\
    & .974 & .920 & .961 & .961 & .949 & 1 & .961 \\ \hline
F2  & .994 & .982 & 1.000 & .998 & .999 & .958 & 1 \\ 
    & .996 & .984 & 1.000 & .999 & .999 & .961 & 1 \\ \hline
\hline
\end{tabular}
\end{minipage}
\hspace{0.5cm}
\begin{minipage}[b]{0.45\linewidth}
\centering
\begin{tabular}{cccccccc}
\hline
\hline
  & EOB & T1   & T2   & T3   & T4   & Et   & F2   \\
\hline
EOB & 1 & .916 & .974 & .938 & .981 & .888 & .970 \\ 
    & 1 & .877 & .973 & .928 & .978 & .841 & .968 \\ \hline
T1  & .916 & 1 & .974 & .926 & .964 & .784 & .975 \\
    & .877 & 1 & .955 & .892 & .947 & .653 & .957 \\ \hline
T2  & .974 & .974 & 1 & .949 & .993 & .861 & .993 \\
    & .973 & .955 & 1 & .932 & .994 & .775 & .995 \\ \hline
T3  & .938 & .926 & .949 & 1 & .943 & .925 & .944 \\
    & .928 & .892 & .932 & 1 & .926 & .876 & .930 \\ \hline
T4  & .981 & .963 & .993 & .943 & 1 & .854 & .995 \\
    & .978 & .947 & .994 & .926 & 1 & .766 & .996\\ \hline
Et  & .888 & .785 & .861 & .925 & .854 & 1 & .852 \\
    & .841 & .653 & .775 & .876 & .767 & 1 & .770 \\ \hline
F2  & .970 & .975 & .993 & .944 & .995 & .853 & 1 \\
    & .968 & .957 & .995 & .930 & .996 & .770 & 1 \\ \hline
\hline
\end{tabular}
\end{minipage}
\end{table*}

For completeness, we also report on the faithfulness of the different 
PN approximants with respect to one another. The faithfulness is the overlap 
between normalized template and signal approximants when maximizing only 
over the time and phase at coalescence, $t_C$ and $\phi_C$. In Tables~
\ref{tab:faithful_I} and \ref{tab:faithful_II}, we list the 
faithfulness for each pair of PN approximants at their highest PN order, that 
is 3.5PN order, except for the EOB model which uses a p4PN order coefficient, 
for both Initial and Advanced LIGO and for each of our reference binaries. 

In the first row and column of the left panel of 
Table \ref{tab:faithful_I}, notice that
every approximant has an overlap of at least 0.97 with the EOB model
for both Initial and Advanced LIGO.  That all approximants have good 
agreement for a low mass binary without searching over mass parameters 
is further evidence that the 3.5PN approximants are rather close 
to one another during the adiabatic inspiral. Note that the T2, T3, T4 and 
F2 approximants all have a faithfulness $\geq 0.99$ with the EOB model, 
while the T1 and Et approximants have somewhat worse agreement at about $0.97$. 
For each pair, the faithfulness for Initial and Advanced LIGO 
are quite similar for these low mass binaries.

In the right panel of Table \ref{tab:faithful_I}, 
we increase the total mass to $10 M_\odot$
while keeping the mass ratio nearly equal. The faithfulness drops for every 
pair of approximants as the merger begins to enter the sensitive band. 
Recall that for these masses, all pairs of approximants can achieve an 
effectualness of at least $0.95$ by searching over the mass parameters. 
When we fix the masses, the T2, T4 and F2 approximants still have very good 
agreement with the EOB model, with faithfulness of $0.97-0.98$. The 
EOB-T3 faithfulness has degraded somewhat to $0.93-0.94$, and the Et and T1 
approximants have rather poor agreement with the EOB model with faithfulness in the 
range $0.84-0.92$. Note that the faithfulness is typically lower for Advanced
LIGO than for Initial LIGO. We attribute this to the signals having a longer 
duration (and thus more time to accumulate a phase difference) in Advanced 
LIGO's wider sensitivity band.

In the left panel of Table \ref{tab:faithful_II}, 
we increase the total mass to $20 M_\odot$
while again keeping the mass ratio nearly equal. Once again, the faithfulness
drops for all cases as the merger and ringdown become more important. The 
T4 and F2 approximants have the best agreement with EOB, they are the only
approximants to achieve an overlap greater than $0.9$ with EOB. The overlap 
between T3 and EOB has dropped dramatically to $0.65$ and $0.72$ for Initial 
and Advanced LIGO respectively.

\begin{table*}
\caption{Same as Table \ref{tab:faithful_I} but for 
$(10.5,9.5)\,M_\odot$ (left panel) and $(10,1.4)\,M_\odot$ 
(right panel) binaries.}
\label{tab:faithful_II}
\vspace{0.5cm}
\begin{minipage}[b]{0.45\linewidth}
\centering
\begin{tabular}{cccccccc}
\hline
\hline
  & EOB & T1   & T2   & T3   & T4   & Et   & F2   \\
\hline
EOB & 1 & .877 & .882 & .650 & .923 & .860 & .910 \\
    & 1 & .811 & .864 & .721 & .910 & .775 & .889 \\ \hline
T1  & .877 & 1 & .972 & .712 & .970 & .817 & .982 \\
    & .811  & 1 & .955 & .785 & .943 & .638 & .966 \\ \hline
T2  & .882 & .972 & 1 & .742 & .968 & .886 & .959 \\
    & .864 & .955 & 1 & .831 & .969 & .784 & .959 \\ \hline
T3  & .650 & .712 & .742 & 1 & .707 & .716 & .709 \\
    & .721 & .785 & .831 & 1 & .794 & .782 & .790 \\ \hline
T4  & .923 & .971 & .968 & .707 & 1 & .906 & .986 \\
    & .910 & .943 & .970 & .794 & 1 & .785 & .988 \\ \hline
Et  & .859 & .817 & .886 & .716 & .906 & 1 & .845 \\
    & .776 & .639 & .784 & .783 & .785 & 1 & .707 \\ \hline
F2  & .909 & .982 & .959 & .708 & .985 & .846 & 1 \\
    & .889 & .967 & .959 & .790 & .988 & .706 & 1 \\ \hline
\hline
\end{tabular}
\end{minipage}
\hspace{0.5cm}
\begin{minipage}[b]{0.45\linewidth}
\centering
\begin{tabular}{cccccccc}
\hline
\hline
  & EOB & T1   & T2   & T3   & T4   & Et   & F2   \\
\hline
EOB & 1 & .977 & .973 & .817 & .859 & .526 & .990 \\
    & 1 & .959 & .972 & .801 & .797 & .413 & .993 \\ \hline
T1  & .977 & 1 & .972 & .796 & .805 & .508 & .991 \\
    & .959 & 1 & .954 & .753 & .691 & .398 & .978 \\ \hline
T2  & .973 & .972 & 1 & .835 & .894 & .543 & .980 \\
    & .972 & .954 & 1 & .820 & .834 & .430 & .976\\ \hline
T3  & .817 & .796 & .835 & 1 & .851 & .778 & .818 \\
    & .801 & .753 & .820 & 1 & .841 & .631 & .798 \\ \hline
T4  & .859 & .805 & .894 & .851 & 1 & .595 & .852 \\
    & .797 & .691 & .834 & .841 & 1 & .456 & .779 \\ \hline
Et  & .526 & .508 & .543 & .778 & .595 & 1 & .525 \\
    & .413 & .398 & .430 & .631 & .456 & 1 & .411 \\ \hline
F2  & .990 & .991 & .980 & .818 & .852 & .525 & 1 \\
    & .993 & .978 & .976 & .799 & .779 & .411 & 1 \\ \hline
\hline
\end{tabular}
\end{minipage}
\end{table*}

The right panel of Table \ref{tab:faithful_II} 
gives the faithfulness for each approximant 
pair for an asymmetric $(10,\,1.4)\,M_\odot$ binary. The EOB-F2 faithfulness is 
very good at $0.99$. The T1 and T2 approximants also have good agreement with 
the EOB model with faithfulness $0.96 - 0.98$. The T3 and T4 have 
poor agreement with the
EOB model with faithfulness $0.80-0.86$. For this mass pair, the Et approximant 
has very poor agreement with all of the others, the faithfulness is  
$\leq 0.60$ for every approximant except T3.

We see a clear trend of decreasing faithfulness as the total mass of the binary 
increases. This is due to the late inspiral, merger and ringdown moving into 
the sensitive band and becoming more important for higher mass binaries. The 
faithfulness is typically lower for Advanced LIGO than Initial LIGO due to its 
broader sensitive band. The faithfulness can vary with mass ratio. For example, for the 
$(10,\,1.4)\,M_\odot$ binary, the T1 and T2 approximants have a better 
faithfulness with the EOB model than the T4 approximant. However, for the nearly equal 
mass binaries, the T4 approximant has the greater faithfulness with the EOB model.
The TaylorF2 approximant is generally the most faithful to the EOB aproximant, 
with one of the highest overlaps in each case. This is another argument for 
using TaylorF2 templates in the mass regime where EOB templates are too 
computationally expensive to be employed in a matched filtering search.

\section{Conclusions}
\label{sec:conclusions}

In this paper we have examined the convergence of the PN
approximation with the view to validating their use in the search for
compact binaries in Initial, Enhanced and Advanced LIGO. We considered seven different 
approximants, each at three different PN orders, a total of 
$21$ waveforms in all. We computed the effectualness of each of the
waveforms with every other at 2PN, 3PN and 3.5PN orders
by using a template bank constructed with a minimal match of $0.99$ 
and Initial and Advanced LIGO noise power spectral densities. 
Our results from a sample of four binaries show that different 
PN approximations are consistent with one another at 
3PN and 3.5PN order.  They begin to differ only when the mass becomes
so large that the plunge phase, not contained in standard PN waveforms 
in the adiabatic approximation, 
enters the detector band. 

\begin{figure*}[t]
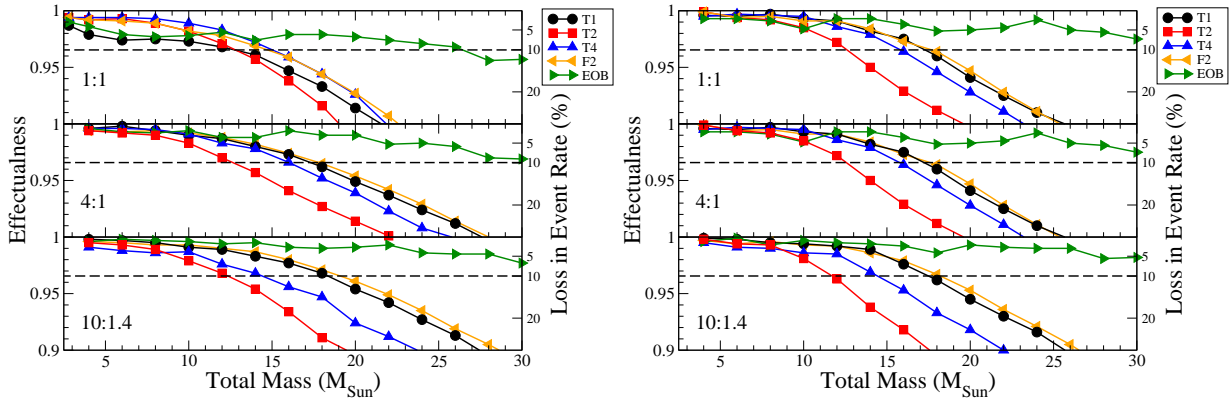

\includegraphics[width=0.45\textwidth]{figures/effectualness-vs-mass-ILIGO.eps}
\includegraphics[width=0.45\textwidth]{figures/effectualness-vs-mass-ALIGO.eps}
\caption{Effectualness (left y-axis) and the corresponding loss in event rate 
(right y-axis) of 3.5PN approximants with the EOB inspiral-merger-ringdown 
signal calibrated to numerical relativity 
in Initial LIGO (left panel) and Advanced LIGO (right panel) 
as a function of total mass for 1:1, 4:1 and 10:1.4 mass ratios. 
The EOB curve is the effectualness between the {\it uncalibrated} 
3.5PN EOB model containing only the inspiral and the {\it calibrated} 
inspiral-merger-ringdown EOB signal. Note that 
any of these approximants are suitable for detection templates below a total 
mass of about $12\,M_\odot$ for both Initial LIGO and Advanced LIGO, 
provided a $10\%$ loss of event rate is deemed acceptable.}
\label{fig:effectualness-vs-mass}
\end{figure*}

The above conclusion is best summarized by Fig. ~\ref{fig:effectualness-vs-mass}, 
where we plot the effectualness of the various PN approximants 
(except for TaylorT3 and TaylorEt that we recommend be discarded, 
since  we have shown that not only do they
 differ considerably from the others but importantly have poorer overlaps with
EOBNR waveforms) with an EOB inspiral-merger-ringdown 
signal as a function of the total mass of the binary. These plots are 
convenient for identifying the $M_{\rm crit}$ above which the PN approximants 
begin to differ with one another. We find that any of the above
 approximants could 
be used as detection templates with less than a $10\%$ loss in event rate 
up to a total mass of $12\,M_\odot$ for both 
Initial and Advanced LIGO. Note that this value of $M_{\rm crit}$  
is limited by the equal-mass case, 
as the value of $M_{\rm crit}$ corresponding to a $10\%$ loss in event rate 
is somewhat greater for mass ratios of 4:1 and 10:1.4. We attribute this result 
to asymmetric binaries accumulating more signal at low frequencies than in the 
equal-mass case. Thus, for a fixed total mass, the merger and ringdown 
are less important for asymmetric binaries than for equal-mass binaries.
Therefore, we conclude that we can safely use any of the above 3.5PN families
as search templates to detect binaries whose total mass is less than
about $12\,M_\odot.$ However, purely from the point of view of 
computational burden TaylorF2 is the least expensive and we recommend
that TaylorF2 at 3.5 PN order be deployed as search templates
below a total mass of $12\,M_\odot$. It is quite remarkable to note that up 
to a total mass of $30 M_\odot$, the {\it uncalibrated} 
EOB model at 3.5PN order is rather close to the calibrated EOB inspiral-merger-ringdown 
signal. In fact, Ref.~\cite{Boyle2008a} found a phase difference 
of {\it only} $0.05$ rads after 30 GW cycles, at roughly 3 GW 
cycles before merger between the EOB at 3.5PN order and the highly accurate 
equal-mass numerical waveform of Caltech/Cornell collaboration.

For systems with total mass larger than about $12\,M_\odot,$ TaylorF2
at 3.5PN might be effectual if the upper cutoff frequency is artificially
extended to a higher frequency. However, this might require a tweaking
of the upper frequency cutoff depending on the noise spectral density
of the detector~\cite{Boyle:2009dg} and the mass ratio of the system, 
and either the extension to unphysical values of $\nu$~\cite{Pan:2007nw,Boyle:2009dg} 
or the inclusion of a p4PN term in the template phase~\cite{Pan:2007nw} calibrated to the 
numerical simulations. We believe that a better alternative for heavier systems are the 
EOB templates calibrated to numerical relativity simulations~\cite{Buonanno-Cook-Pretorius:2007,Pan:2007nw,Damour:2007xr,Buonanno:2007pf,Damour:2007yf,
Damour:2007vq,Damour:2008te,Boyle2008a,Damour:2009b,Buonanno:2009qa}. 
The most recent EOB models are in near perfect agreement with the most accurate numerical simulations to date, although
only a small number of systems corresponding to different mass ratios
have been studied so far. Nevertheless, a physical model with physically 
meaningful parameters is a far safer bet as search templates unless,
of course, if the model in question is not in agreement with the waveform
predicted by numerical relativity. So far, the EOB is the best physical
model we have and this is what we recommend be used to search for binaries
with masses greater than about $12\,M_\odot.$

In this paper we adopted the preliminary, fiducial EOB model of Ref.~\cite{Buonanno:2007pf}, 
because it is the EOB model currently available in LAL and it is used 
for searches by  Initial LIGO. For completeness, here we quantify the closeness 
between the EOB model used in this paper and a most recent 
improved version of the EOB model~\cite{Buonanno:2009qa} (which is similar to the 
one of Ref.~\cite{Damour:2009b}). The latter 
was calibrated to longer and more accurate numerical waveforms 
generated by the Caltech/Cornell pseudo-spectral code~\cite{Scheel:2008rj}. 
Reference~\cite{Buonanno:2009qa} found that the faithfulness of the 
improved EOB model to these highly accurate numerical waveforms 
is better than 0.999. In Table\ \ref{tab:fEOBcomparison}, we show 
both the faithfulness and the effectualness of the EOB model~\cite{Buonanno:2007pf} 
to the improved EOB model~\cite{Buonanno:2009qa} using noise
spectral densities of Initial LIGO, as well as the bias in the parameters 
$M$ and $\nu$ when achieving the effectualness. The search for
effectualness in this test is done continuously in the parameter space,
instead of using a template bank. Although there is some systematic
trend in the numbers due to the difference in the EOB models, 
the main result is that the faithfulness and the effectualness are
always better than 0.97 and 0.995, respectively. Assuming the numerical
waveforms of Ref.~\cite{Buonanno:2009qa} are exact, the EOB model of Ref.~\cite{Buonanno:2007pf} 
used in this paper is accurate for detection purpose with a loss of event rates of $\sim 10 \%$, 
and may cause $\sim 10\%$ bias in estimating the mass parameters.
\begin{table}
\caption{Effectualness and faithfulness of the EOB fiducial model~\cite{Buonanno:2007pf} 
used in this paper (and currently employed by Initial LIGO) to the most recently improved 
EOB model~\cite{Buonanno:2009qa}. We also show the bias in the parameters 
$M$ and $\nu$ when achieving the effectualness. For each pair, the top
 number is Initial LIGO while the bottom number is Advanced LIGO. 
The sign of the bias is such that in all cases the fiducial EOB templates 
slightly overestimate the total mass $M$ and underestimate the mass ratio 
$\nu$ of the improved EOB signal.
}
\label{tab:fEOBcomparison}

\begin{tabular}{|c|ccc|c|}
\hline
& Effectualness & $\Delta M/M$ & $\Delta\nu/\nu$ & Faithfulness
\\ \hline
\multirow{1}{*}{$(1.4,1.4)M_\odot$} 
& 0.999 & 0.98\% & -1.63\% & 0.992 \\
& 0.999 & 0.98\% & -1.63\% & 0.995 \\ 
\hline
\multirow{1}{*}{$(1.38,1.42)M_\odot$} 
& 0.999 & 0.96\% & -1.60\% & 0.992 \\
& 0.999 & 0.89\% & -1.49\% & 0.995 \\ 
\hline
\multirow{1}{*}{$(5,5)M_\odot$}  
& 0.997 & 1.32\% & -2.12\% & 0.973 \\
& 0.999 & 2.06\% & -3.47\% & 0.976 \\ 
\hline
\multirow{1}{*}{$(4.8,5.2)M_\odot$} 
& 0.999 & 2.42\% & -4.08\% & 0.973 \\
& 0.999 & 2.11\% & -3.54\% & 0.976 \\ 
\hline
\multirow{1}{*}{$(10,10)M_\odot$} 
& 0.999 & 2.70\% & -4.62\% & 0.974 \\
& 0.999 & 2.59\% & -4.39\% & 0.962 \\ 
\hline
\multirow{1}{*}{$(9.5,10.5)M_\odot$} 
& 0.998 & 1.40\% & -1.94\% & 0.974 \\
& 0.997 & 2.67\% & -4.54\% & 0.964 \\ 
\hline
\multirow{1}{*}{$(15,15)M_\odot$} 
& 0.995 & 4.80\% & -9.98\% & 0.987 \\
& 0.999 & 2.49\% & -4.23\% & 0.973 \\ 
\hline
\multirow{1}{*}{$(25,25)M_\odot$} 
& 0.995 & 4.95\% & -12.6\% & 0.982 \\
& 0.994 & 3.00\% & -5.56\% & 0.985 \\ 
\hline
\end{tabular}
\end{table}

In this study we considered PN waveforms in the so-called 
restricted PN approximation. Restricted waveforms contain only the
second harmonic of the orbital frequency. Inclusion of other harmonics
is necessary, especially when a binary is arbitrarily oriented with
respect to a detector and the component masses are dissimilar. Recent
studies~\cite{VanDenBroeck:2006ar,VanDenBroeck:2006qu} have shown 
the tremendous advantage of including these other harmonics in the
GW templates.
Therefore, it is necessary that a future effort undertakes a study similar
to this, but includes all the amplitude corrections. Furthermore,  
Ref.~\cite{Damour:2009a} has shown that by supplementing the PN results
by the available test particle results up to 5.5PN improves the
match between the EOB models and numerical relativity simulations.
This can be expected to lead to  further improvements in the results 
obtained here in the future.

\section*{Acknowledgements}
We thank Luc Blanchet for useful comments and Steve Fairhurst for dicussions. 
We thank Michele Vallisneri for carefully reading the paper.
B.~R.~Iyer thanks Cardiff university for hospitality and support during  
this work. 
A.B., E.O. and Y.P. acknowledge support from NSF Grant No. PHY-0603762.

\bibliography{references}

\end{document}